\documentclass[10pt,journal,compsoc]{IEEEtran}
\usepackage{amsmath}
\usepackage{lineno,hyperref}
\usepackage{float} 
\usepackage{xcolor,soul,framed} 
\usepackage{tabularx}
\usepackage{dblfloatfix}
\usepackage{tabulary}
\usepackage{pgffor}
\usepackage{subfig}
\usepackage{lipsum}
\usepackage{pdflscape}
\usepackage{adjustbox}

\usepackage{bbding}
\usepackage{pifont}
\usepackage{wasysym}
\usepackage{amssymb}
\usepackage{hyperref}

\usepackage{ragged2e}
\usepackage{booktabs}

\usepackage{tabularx,colortbl}

\usepackage{tikz}
\usetikzlibrary{patterns} 
\usepackage{collcell}
\usepackage{diagbox}
\usepackage{rotating}
\usepackage{arydshln}
\usepackage{multirow}

\usepackage{cuted}
\usepackage{flushend}

\usepackage{algorithm}
\usepackage{algpseudocode}
\usepackage{parskip}

\usepackage[numbers]{natbib}

\bstctlcite{IEEEexample:BSTcontrol}
\usepackage{amsthm}

\newtheorem{definition}{Definition}[section]

\newtheorem{problem}{Problem}

\newcommand{\cmark}{\ding{51}}
\newcommand{\xmark}{\ding{55}}

\begin{document}
%
\title{AlertStar: Path-Aware Alert Prediction on 
       Hyper-Relational Knowledge Graphs}

\author{Zahra~Makki Nayeri,
        Mohsen~Rezvani
\IEEEcompsocitemizethanks{\IEEEcompsocthanksitem Zahra Makki Nayeri, Faculty
of Computer Engineering at Shahrood University of Technology, Shahrood, Iran. E-mail: zmakki@shahroodut.ac.ir

\IEEEcompsocthanksitem Mohsen Rezvani, Faculty
of Computer Engineering at Shahrood University of Technology, Shahrood, Iran.
E-mail: mrezvani@shahroodut.ac.ir
}
}
\IEEEtitleabstractindextext{%
\justify{

\begin{abstract}
Cyber-attacks continue to grow in scale and sophistication, yet existing network intrusion detection approaches lack the semantic depth required for path reasoning over attacker--victim interactions. We address this by first modelling network alerts as a knowledge graph, then formulating hyper-relational alert prediction as a hyper-relational knowledge graph completion~(HR-KGC) problem, representing each network alert as a qualified statement $(h, r, t, \mathcal{Q})$, where $h$ and $t$ are source and destination IPs, $r$ denotes the attack type, and $\mathcal{Q}$ encodes flow-level metadata such as timestamps, ports, protocols, and attack intensity, going beyond standard KGC binary triples $(h, r, t)$ that would discard this contextual richness. We introduce five models across three contributions: first, Hyper-relational Neural Bellman--Ford (HR-NBFNet) extends Neural Bellman--Ford Networks to the hyper-relational setting with qualifier-aware multi-hop path reasoning, while its multi-task variant MT-HR-NBFNet jointly predicts tail, relation, and qualifier-value within a single traversal pass; second, AlertStar fuses qualifier context and structural path information entirely in embedding space via cross-attention and learned path composition, and its multi-task extension MT-AlertStar eliminates the overhead of full knowledge graph propagation; third, HR-NBFNet-CQ extends qualifier-aware representations to answer complex first-order logic queries, including one-hop, two-hop chain, two-anchor intersection, and union, enabling multi-condition threat reasoning over the alert knowledge graph. Evaluated inductively on the Warden and UNSW-NB15 benchmarks across three qualifier-density regimes, AlertStar and MT-AlertStar achieve superior MR, MRR, and Hits@k, demonstrating that local qualifier fusion is both sufficient and more efficient than global path propagation for hyper-relational alert prediction.
\end{abstract}}

}

\maketitle

\IEEEdisplaynontitleabstractindextext

\IEEEpeerreviewmaketitle

\IEEEraisesectionheading{\section{Introduction}
\label{sec:introduction}}

Cyber-attacks continue to grow in scale and
sophistication, rendering purely reactive defences
insufficient for modern network security.
Operational intrusion detection systems~(IDSs) are
highly sensitive by design, routinely generating
tens of thousands of alerts per day~\cite{husak2021predictive},
yet the most dangerous threats --- advanced persistent
threat~(APT) campaigns --- are rarely revealed by
any single alert. APT actors deliberately decompose
their campaigns into individually innocuous steps:
reconnaissance, initial compromise, lateral movement,
and data exfiltration, each generating only a
low-priority alert that may be dismissed in
isolation~\cite{husak2021predictive}.

The central challenge facing Security Operations
Centers~(SOCs) and Computer Emergency Response
Teams~(CERTs) is reasoning \emph{across} sequences
of alerts distributed over time and across different
host pairs to reconstruct the underlying campaign \cite{9684707}.
Consider a representative scenario: a denial-of-service
alert at 02:14, a reconnaissance alert on a different
source IP at 03:47, and a data exfiltration alert on a
third IP pair at 06:22 --- each assessed as medium
priority in isolation, yet collectively the signature
of a coordinated lateral-movement campaign. A CERT
team that could have predicted the second and third
alerts from the first could have intervened before
exfiltration. This is the problem of \emph{alert
prediction}: given the observed alert history, infer
which IP will be targeted next, under which attack
type, and via which propagation path.

Knowledge graphs~(KGs) offer a principled solution
by representing attacker--victim interactions as
relational triples $(h, r, t)$, enabling heterogeneous,
semantically rich reasoning over unseen entities ---
a critical capability for inductive alert prediction.
However, standard KGC models are limited to binary
triples, which cannot encode the auxiliary flow-level
metadata that distinguishes network alerts of the
same type. Two alerts sharing the same source IP,
attack type, and destination IP may represent
fundamentally different threat scenarios depending
on their timestamp, port, protocol, and flow
intensity. This motivates \emph{hyper-relational}
knowledge graphs, where each alert is represented
as a qualified statement $(h, r, t, \mathcal{Q})$
with $\mathcal{Q}$ encoding auxiliary key--value
pairs beyond the core triple.

Existing hyper-relational KGC models~(StarE~\cite{galkin2020message},
StarQE~\cite{alivanistos2021query}, HyNT~\cite{chung2023representation}) are
predominantly transductive, limiting their
applicability when new IP addresses unseen during
training must be reasoned about at inference time.
Neural Bellman--Ford Networks~(NBFNet)~\cite{zhu2021neural}
address inductive link prediction via path-based
message passing, but operate on binary triples with
no mechanism for qualifier context.

In this work, we model network alerts as a
hyper-relational knowledge graph and formulate alert
prediction as an HR-KGC problem. We introduce five
models across three paradigms: qualifier-aware
path propagation, embedding-based fusion, and
complex query answering over alert graphs.

\noindent The main contributions are:
\begin{itemize}
  \item \textbf{HR-KGC alert benchmarks.}
    We formulate alert prediction as HR-KGC and
    construct inductive benchmarks from Warden and
    UNSW-NB15 across three qualifier-density regimes
    (Q33, Q66, Q100), retaining 33\%, 66\%,
and 100\% of qualifier pairs respectively, to study robustness to metadata
    availability.

  \item \textbf{HR-NBFNet and MT-HR-NBFNet.}
    HR-NBFNet extends NBFNet to the hyper-relational
    setting by injecting qualifier context at every
    message-passing step. MT-HR-NBFNet augments this
    backbone with joint prediction of tail, relation,
    and qualifier-value objectives within a single
    Bellman--Ford pass at no additional graph-traversal
    cost.

  \item \textbf{AlertStar and MT-AlertStar.}
    AlertStar fuses qualifier-aware multi-head
    cross-attention with a learned path-composition
    branch via a trainable scalar gate, operating
    entirely in embedding space without graph traversal,
    yielding more stable optimisation on sparse,
    inductive alert graphs than HR-NBFNet.
    MT-AlertStar extends this with joint prediction
    from a shared Transformer encoder, completing a
    symmetric multi-task design across  paradigms
    with stronger qualifier supervision than
    graph-propagation alternatives.

  \item \textbf{HR-NBFNet-CQ.}
    HR-NBFNet-CQ extends qualifier-aware representations
    to answer four first-order logic query types ---
    one-hop~(1p), two-hop chain~(2p), two-anchor
    intersection~(2i), and two-anchor union~(2u) --- enabling
    complex multi-condition threat investigations
    directly over the alert knowledge graph.

  \item \textbf{Comprehensive evaluation.}
    We evaluate all five models against state-of-the-art
    HR-KGC baselines under inductive settings, with
    complexity analysis showing AlertStar achieves up
    to $50\times$ per-epoch speedup over HR-NBFNet,
    and four ablation studies~(A1--A4) dissecting
    component contributions and qualifier-density
    sensitivity.
\end{itemize}

The remainder of this paper is organised as follows.
Section~\ref{sec:preliminaries} introduces background
on hyper-relational KGs and the threat model.
Section~\ref{sec:models} formalises the problem and
presents the five models.
Section~\ref{sec:experiments} describes experiments
and evaluation.
Section~\ref{sec:related} reviews related work.
Section~\ref{sec:discussion} analyses results.
Section~\ref{sec:conclusion} concludes and outlines
future work.

\section{Preliminaries and Threat Model}
\label{sec:preliminaries}

This section introduces the formal background required for the
remainder of the paper, and defines the operational threat model
and assumptions under which the proposed models are evaluated.

\subsection{Knowledge Graphs}
\label{ssec:kg}

\begin{definition}[Knowledge Graph]
\label{def:kg}
A knowledge graph is a directed, labeled multigraph
$\mathcal{G} = (\mathcal{E}, \mathcal{R}, \mathcal{T})$,
where $\mathcal{E}$ is a finite set of entities,
$\mathcal{R}$ is a finite set of relation types, and
$\mathcal{T} \subseteq \mathcal{E} \times \mathcal{R}
\times \mathcal{E}$ is a set of facts, each represented
as a triple $(h, r, t)$ with head entity $h \in \mathcal{E}$,
relation $r \in \mathcal{R}$, and tail entity $t \in \mathcal{E}$.
\end{definition}

In the network alert setting, $\mathcal{E}$ is the set of
observed IP addresses, $\mathcal{R}$ is the set of attack
categories (e.g., Recon, DoS, DDoS, Anomaly), and each triple
$(h, r, t) \in \mathcal{T}$ records that source IP~$h$ launched
an attack of type~$r$ against destination IP~$t$.

\begin{definition}[Knowledge Graph Completion]
\label{def:kgc}
Given a knowledge graph $\mathcal{G}$, the knowledge graph
completion (KGC) task is to score and rank candidate tail
entities $t'$ for an incomplete query $(h, r, ?)$, or
equivalently candidate head entities $h'$ for a query
$(?, r, t)$, using a scoring function
$f: \mathcal{E} \times \mathcal{R} \times \mathcal{E}
\rightarrow \mathbb{R}$.
\end{definition}

Applied to alert prediction, the KGC task answers: given that
source IP~$h$ is conducting an attack of type~$r$, which
destination IP~$t'$ is the most likely next target.

\subsection{Hyper-Relational Knowledge Graphs}
\label{ssec:hrkg}
Standard KGs represent each alert as a binary triple $(h, r, t)$,
discarding the flow-level metadata that accompanies every alert
record. Hyper-relational knowledge graphs (HR-KGs) extend this
representation to incorporate predicate-scoped contextual
attributes known as \emph{qualifiers}~\cite{galkin2020message}.

\begin{definition}[Hyper-Relational Knowledge Graph]
\label{def:hrkg}
A hyper-relational knowledge graph is a set of qualified
statements $\mathcal{G}^+ = \{(h, r, t, \mathcal{Q})\}$,
where $(h, r, t) \in \mathcal{E} \times \mathcal{R}
\times \mathcal{E}$ is the main triple and
$\mathcal{Q} = \{(q_k^i, q_v^i)\}_{i=1}^{n}$
is a set of qualifier pairs with
$q_k^i \in \mathcal{R}$ (qualifier relations) and
$q_v^i \in \mathcal{E}$ (qualifier entities).
\end{definition}

\paragraph*{StarE}
StarE~\cite{galkin2020message} is a message-passing GNN
designed for representation learning over HR-KGs.
It augments each main relation embedding with a
qualifier vector computed by composing and aggregating
all qualifier pairs of a given fact:
\begin{equation}
  \mathbf{h}_q = \mathbf{W}_q
    \sum_{(q_k,\,q_v)\in\mathcal{Q}}
    \phi_q(\mathbf{h}_{q_k},\,\mathbf{h}_{q_v}),
  \label{eq:stare_qual}
\end{equation}
where $\mathbf{W}_q$ is a learned projection matrix and
$\phi_q$ is a composition function such as DistMult or
RotatE~\cite{sun2019rotate}.
The qualifier vector is then merged with the main relation
embedding $\mathbf{h}_r$ via:
\begin{equation}
  \gamma(\mathbf{h}_r,\mathbf{h}_q)
    = \alpha \odot \mathbf{h}_r
    + (1-\alpha) \odot \mathbf{h}_q,
  \label{eq:stare_gamma}
\end{equation}
where $\alpha \in [0,1]$ is a hyperparameter controlling
the flow of qualifier information into the relation
representation. Node embeddings are then updated by
message passing:
\begin{equation}
  \mathbf{h}_v = f\!\left(
    \sum_{(u,r)\in\mathcal{N}(v)}
    \mathbf{W}_{\lambda(r)}\,
    \phi_r\!\left(\mathbf{h}_u,\,
    \gamma(\mathbf{h}_r,\mathbf{h}_q)_{vu}\right)
  \right),
  \label{eq:stare_update}
\end{equation}
where $\phi_r$ is a relation composition function,
$\mathbf{W}_{\lambda(r)}$ is a direction-specific weight
matrix for incoming, outgoing, and self-loop edges, and
$\mathcal{N}(v)$ denotes the set of incoming edges to $v$.

\subsection{Neural Bellman-Ford Networks}
\label{ssec:nbfnet}

Path-based knowledge graph completion methods reason over multi-hop paths between
a query source entity and candidate answer entities.
Neural Bellman-Ford Networks (NBFNet)~\cite{zhu2021neural} provide a principled
and general framework for this by casting link prediction as a \emph{pair
representation learning} problem.

\begin{definition}[Pair Representation]
\label{def:pair_repr}
For a query $(u, q, ?)$, the pair representation $\mathbf{h}_q(u,v)$ is defined
as the \emph{generalised sum} of all path representations between $u$ and $v$,
where each path representation is the \emph{generalised product} of the edge
representations along the path:

\begin{equation}
  \mathbf{h}_q(u,v)
    = \bigoplus_{P \in \mathcal{P}_{uv}} \mathbf{h}_q(P),
  \label{eq:pair_repr}
\end{equation}
\begin{equation}
  \mathbf{h}_q(P = (e_1,\ldots,e_{|P|}))
    = \bigotimes_{i=1}^{|P|} \mathbf{w}_q(e_i).
  \label{eq:path_repr}
\end{equation}

where $\oplus$ and $\otimes$ denote a commutative summation operator and a
(not necessarily commutative) multiplication operator, respectively, and
$\mathbf{w}_q(e)$ is the representation of edge $e$ under query relation~$q$.
\end{definition}

\noindent
This formulation subsumes several classical link-prediction heuristics as special
cases: Katz index uses $\oplus{=}+$ and $\otimes{=}{\times}$; graph distance uses
$\oplus{=}\min$ and $\otimes{=}+$; widest path uses $\oplus{=}\max$ and
$\otimes{=}\min$~\cite{zhu2021neural}.

\paragraph*{Generalised Bellman-Ford Algorithm.}
Directly computing Equations~\eqref{eq:pair_repr}--\eqref{eq:path_repr} is
intractable, because the number of paths grows exponentially with path length.
When $\langle\oplus,\otimes\rangle$ satisfy a \emph{semiring}~\cite{zhu2021neural},
the pair representation can instead be computed efficiently via the generalised
Bellman-Ford algorithm.
Letting $\mathbb{1}_q$ and $\mathbb{0}_q$ denote the multiplication and summation
identities of the semiring, the algorithm initialises
\begin{equation}
  \mathbf{h}_q^{(0)}(u,v)
    \leftarrow \mathbb{1}_q(u = v),
  \label{eq:nbfnet_init}
\end{equation}
and iterates

\begin{equation}
\begin{split}
  \mathbf{h}_q^{(t)}(u,v)
    &\leftarrow
    \left(
      \bigoplus_{(x,r,v)\in\mathcal{E}(v)}
      \mathbf{h}_q^{(t-1)}(u,x) \otimes \mathbf{w}_q(x,r,v)
    \right) \\
    &\quad \oplus\, \mathbf{h}_q^{(0)}(u,v),
\end{split}
  \label{eq:nbfnet_prop}
\end{equation}

where the residual connection to $\mathbf{h}_q^{(0)}(u,v)$ preserves the boundary
condition at every layer.
The algorithm computes $\mathbf{h}_q(u,v)$ for a fixed source $u$, a fixed query
relation $q$, and \emph{all} $v\in\mathcal{V}$ in parallel, exploiting the
distributive property of $\otimes$ over $\oplus$ to avoid explicit path enumeration.
Setting $\oplus{=}\min$ and $\otimes{=}+$ recovers the original Bellman-Ford
shortest-path algorithm~\cite{bellman1958routing}.

\paragraph*{Neural Bellman-Ford Networks}
NBFNet relaxes the semiring assumption and replaces the hand-crafted operators
with three learned neural components:

\begin{enumerate}
  \item \textbf{INDICATOR} replaces the boundary condition
        $\mathbb{1}_q(u{=}v)$.  In practice it is instantiated as
        $\mathbf{1}(u{=}v)\cdot\mathbf{e}_q$, where $\mathbf{e}_q\in\mathbb{R}^d$
        is a learned query-relation embedding.
  \item \textbf{MESSAGE} replaces the binary multiplication operator $\otimes$.
        It is instantiated using relational operators from knowledge-graph
        embeddings~\cite{sun2019rotate},
        e.g.\ translation ($\mathbf{h}+\mathbf{w}_q$) or element-wise
        multiplication ($\mathbf{h}\odot\mathbf{w}_q$).
  \item \textbf{AGGREGATE} replaces the $n$-ary summation operator $\bigoplus$.
        It is a permutation-invariant set aggregation
        function, e.g.\ sum, mean, max, or PNA, followed
        by a linear transformation and a non-linear activation.
\end{enumerate}

\noindent
Substituting these functions into
Equations~\eqref{eq:nbfnet_init}--\eqref{eq:nbfnet_prop} gives the NBFNet update,
where $(x,r,v)\in\mathcal{E}(v)$ denotes the set of incoming
edges to node $v$:
\begin{align}
  \mathbf{h}_v^{(0)}
    &\leftarrow \textsc{Indicator}(u,v,q),
  \label{eq:nbfnet_ind} \\
  \mathbf{h}_v^{(t)}
    &\leftarrow \textsc{Aggregate}^{(t)}\!\left(
        \left\{
          \textsc{Message}^{(t)}\!\left(\mathbf{h}_x^{(t-1)},\,
          \mathbf{w}_q(x,r,v)\right)
        \right\} \right.\nonumber\\
    &\qquad\qquad \left. \cup\,
        \left\{\mathbf{h}_v^{(0)}\right\}
      \right). \nonumber
  \label{eq:nbfnet_update}
\end{align}

After $L$ propagation steps the pair representation $\mathbf{h}_v^{(L)}$ is
scored by a shallow MLP:
\begin{equation}
  f(u,q,v) = \sigma\!\left(\textsc{MLP}\!\left(\mathbf{h}_v^{(L)}\right)\right).
  \label{eq:nbfnet_score}
\end{equation}

\subsection{Threat Model and Assumptions}
\label{ssec:threat_model}
We adopt a bounded adversarial model tailored to the
properties of the evaluated datasets. The following
assumptions define the operational threat context.

\textbf{A1 --- External and Internal Adversaries.}
The adversary may originate from outside or within the
network perimeter. Both attacker positions are handled
uniformly: source and destination IPs are entities in
$\mathcal{E}$ regardless of their network location.

\textbf{A2 --- Limited-Knowledge Adversary.}
The adversary has no knowledge of the alert prediction
model, its parameters, or its outputs, ensuring that
observed patterns reflect genuine attacker behavior~\cite{9684707}.

\textbf{A3 --- Unidirectional Attack Interactions.}
Each alert represents a unidirectional flow from a source
(attacker) IP to a destination (victim) IP. A victim may
subsequently launch a counter-attack toward the original
attacker, but such interactions are captured as separate
alert events in a different time window, rather than
modeled as a single simultaneous bidirectional exchange \cite{xiong2023shrinking}.

\textbf{A4 --- Role Disjointness per Interaction.}
Within a single alert, an IP plays exactly one role:
attacker ($h$) or victim ($t$). An IP may switch roles
across alerts but not within the same event.

\textbf{A5 --- Closed-World Attack Taxonomy.}
Predicted attack categories are limited to those annotated
in the training data. While our inductive approach
generalizes to unseen entities at test time, it can only
predict whether an interaction exists between previously
unseen IPs --- it cannot infer the attack type for
categories not observed during training. Zero-day or
novel attack taxonomies therefore remain outside the
current scope.

\section{Methodology}
\label{sec:models}

\begin{figure}[t]
    \centering
    \includegraphics[width=\linewidth]{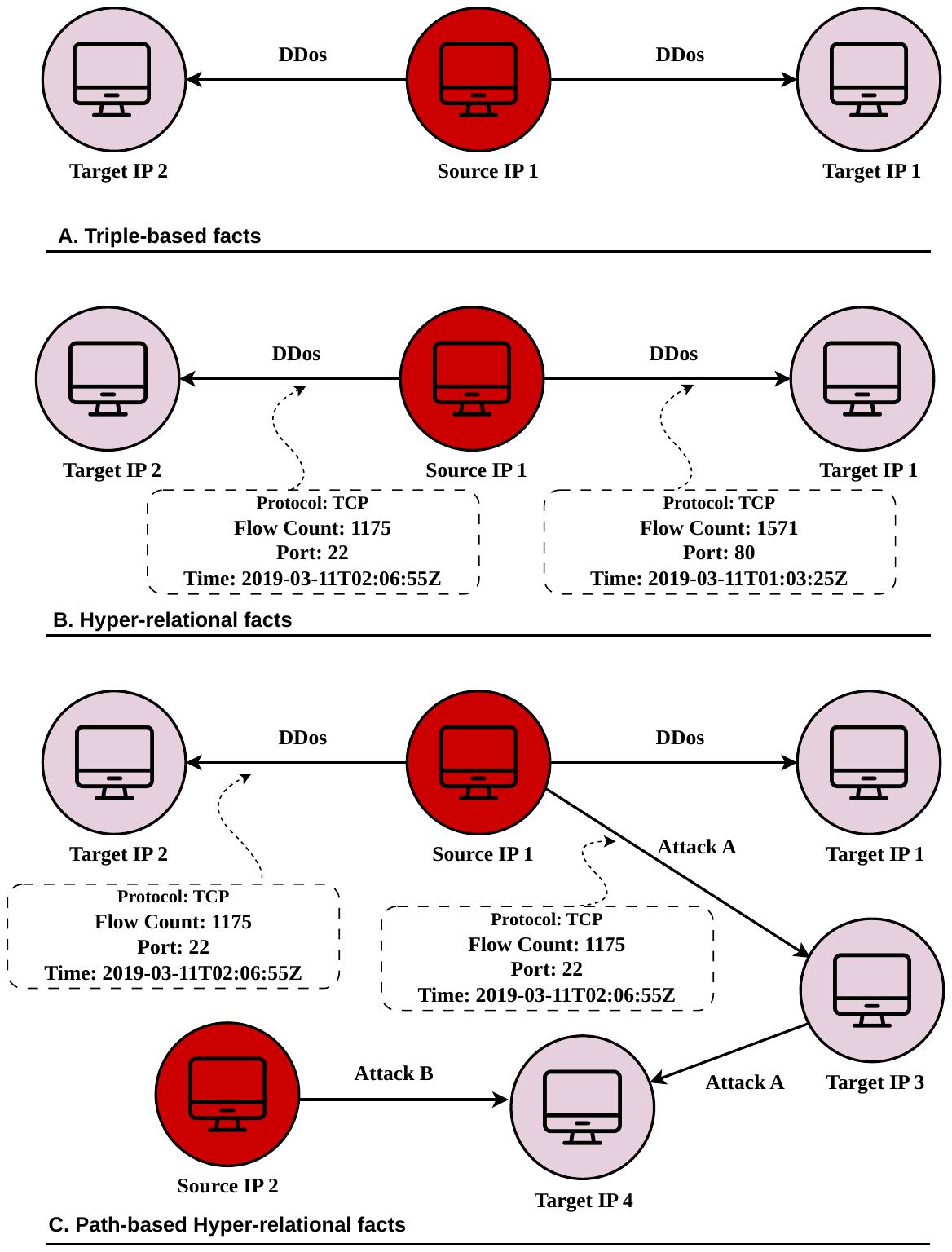}
    \caption{Three representational levels: (A)
    triple-based facts, (B)
    hyper-relational facts, (C) path-based
    hyper-relational reasoning.}
\label{fig:kg_progression}
\end{figure}
Figure~\ref{fig:kg_progression} illustrates the
three representational levels motivating this work.
Triple-based facts~(A) cannot distinguish two
attacks sharing the same $(h,r,t)$ structure.
Hyper-relational facts~(B) solve this ambiguity
by attaching flow-level qualifiers, protocol,
port, flow count, and detect time, to each
alert. Path-based hyper-relational reasoning~(C)
further captures multi-hop attacker pivoting that
qualifier-enriched node representations alone
cannot detect.
These three levels motivate a three-paradigm
investigation.

\textbf{Paradigm~I: Graph Propagation.}
\textbf{HR-NBFNet} extends NBFNet~\cite{zhu2021neural}
to the hyper-relational setting by injecting
qualifier context at every message-passing step
via per-edge DistMult embeddings.
\textbf{MT-HR-NBFNet} augments this backbone with
three auxiliary tasks, tail, attack type, and
qualifier-value prediction, sharing a single
Bellman--Ford pass at no additional graph-traversal
cost. However, Bellman--Ford propagation has a
known scalability limitation:
$\mathcal{O}(|\mathcal{T}|\,Q_{\max}\,d)$ per
query, with path counts growing exponentially with
graph diameter. In dense alert graphs where a
single attacker IP interacts with hundreds of
victims, this overhead becomes prohibitive and
redundant traversals beyond moderate path lengths
carry negligible signal~\cite{zhu2021neural}.

\textbf{Paradigm~II: Embedding-Based.}
\textbf{AlertStar} fuses qualifier context and
structural path information entirely in embedding
space via qualifier-aware multi-head
cross-attention and a feed-forward
path-composition branch, balanced by a trainable
scalar gate, no graph traversal is performed.
\textbf{MT-AlertStar} extends this with a
Transformer encoder and the same three auxiliary
tasks, completing a symmetric multi-task design
across paradigms. The empirical comparison
between HR-NBFNet and AlertStar directly answers
whether global path reasoning is worth its
computational cost for alert prediction.

\textbf{Paradigm III: HR-NBFNet-CQ.} Answering
complex first-order logic queries, one-hop,
two-hop chain, two-anchor intersection, and two-anchor
union, over the hyper-relational alert graph,
enabling multi-condition threat investigations
by SOC and CERT analysts.

\subsection{Alert Graph Representation}
\label{ssec:alert_model}
Table~\ref{tab:warden_sample} illustrates the three
representational levels applied to alert data.
Each alert is first modeled as a binary triple
$(h, r, t)$ --- source IP, attack category, and
destination IP --- forming $\mathcal{G} =
(\mathcal{E}, \mathcal{R}, \mathcal{T})$. This
captures who attacked whom, but treats all alerts
of the same category between the same IP pair as
equivalent, discarding flow-level context.
To resolve this ambiguity, we extend to a
hyper-relational KG $\mathcal{G}^+$ by attaching
qualifier pairs $\mathcal{Q}$ encoding detect time,
flow count, port, and protocol to each triple ---
for example, distinguishing the two DDoS alerts in
Figure~\ref{fig:kg_progression} by port and flow count.
The StarE encoder~\cite{galkin2020message} incorporates
this context into message passing via
Eqs.~\eqref{eq:stare_qual}--\eqref{eq:stare_update}.
However, qualifier-enriched node representations
cannot detect how an attacker pivots across multiple
victims via intermediate nodes
(Figure~\ref{fig:kg_progression}(C)).
NBFNet~\cite{zhu2021neural} addresses this by
computing a pair representation $\mathbf{h}_q(u,v)$
for each source--target pair via multi-hop path
propagation
(Eqs.~\eqref{eq:nbfnet_init}--\eqref{eq:nbfnet_score}),
enabling detection of coordinated multi-stage attacks
and lateral movement patterns beyond what local
neighbourhood aggregation can capture.

\begin{table}[t]
\centering
\caption{Example alert records and their
qualified-statement representation
$(h, r, t, \mathcal{Q})$. The qualifier set
$\mathcal{Q}$ encodes detect time, flow count,
port, and protocol.}
\label{tab:warden_sample}
\renewcommand{\arraystretch}{1.2}
\resizebox{\columnwidth}{!}{%
\begin{tabular}{llllllll}
\toprule
\textbf{Detect Time} & \textbf{Flow Count} &
\textbf{Source IP} $(h)$ & \textbf{Target IP} $(t)$ &
\textbf{Port} & \textbf{Protocol} &
\textbf{Category} $(r)$ \\
\midrule
2019-03-11 00:05 & 17{,}094 & 185.192.59.136
  & 142.252.135.136 & 22  & TCP & Recon Scan \\
2019-03-12 00:45 & 5{,}113  & 78.234.46.141
  & 142.252.32.63   & 443 & TCP & Availability DoS \\
2019-03-14 00:25 & 15       & 185.192.59.136
  & 142.252.32.63   & 81  & TCP & Availability DDoS \\
2019-03-14 00:25 & 39       & 78.234.46.141
  & 142.252.32.63   & 22  & UDP & Anomaly Traffic \\
\midrule
\multicolumn{7}{l}{\textit{Qualified-statement
  mapping for row~1:}} \\
\multicolumn{7}{l}{$h$ = 185.192.59.136,\quad
  $r$ = Recon Scan,\quad
  $t$ = 142.252.135.136} \\
\multicolumn{7}{l}{$\mathcal{Q}$ =
  \{detectTime: 2019-03-11 00:05,\
  flowCount: 17094,\ port: 22,\
  protocol: TCP\}} \\
\bottomrule
\end{tabular}
}
\end{table}

\subsection{Paradigm~I: Hyper-Relational NBFNet}
\label{ssec:hrnbfnet}

HR-NBFNet extends NBFNet~\cite{zhu2021neural} to the
hyper-relational setting by injecting qualifier context
at every propagation step, in the query-conditioned
initialisation and in each per-edge message. Notation and base formulation follow Section~\ref{sec:preliminaries}.

\begin{figure}[t]
    \centering
    \includegraphics[width=\linewidth]{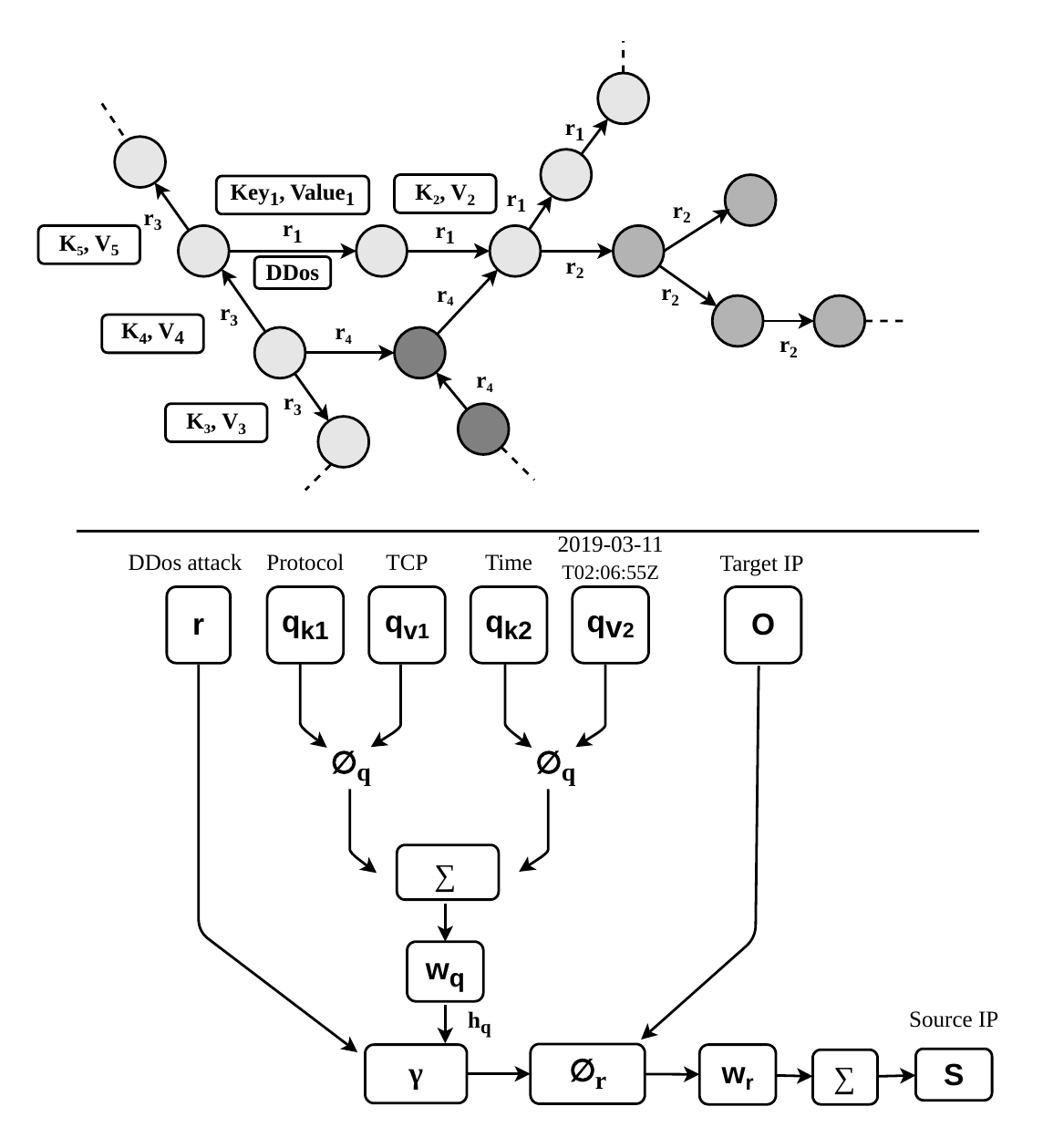}
    \caption{HR-NBFNet combines StarE qualifier
encoding~\cite{galkin2020message} with NBFNet
Bellman-Ford propagation~\cite{zhu2021neural}.
Qualifier pairs are composed via $\phi_q$,
aggregated, projected by $\mathbf{W}_q$, and
merged with $\mathbf{h}_r$ via $\gamma$, injecting
flow-level context into the pair representation
$\mathbf{h}_{uvqq'}^{(t)}$ at every layer.
Inverse edges $(t, r^{-1}, h)$ enable bidirectional
propagation for both tail and head prediction.}
    \label{fig:hrnbfnet}
\end{figure}

\noindent\textbf{Qualifier vector.}
For each edge $(u, r, v)$, the qualifier vector is:
\begin{equation}
  \mathbf{h}_q = \mathbf{W}_q
    \sum_{(q_k, q_v)\in\mathcal{Q}_{r_{vu}}}
    \mathbf{h}_{q_k} \odot \mathbf{h}_{q_v}.
  \label{eq:hr_hq}
\end{equation}

\noindent\textbf{Relation--qualifier merging.}
\begin{equation}
  \gamma(\mathbf{h}_r, \mathbf{h}_q)
    = \alpha \odot \mathbf{h}_r
    + (1{-}\alpha) \odot \mathbf{h}_q,
  \label{eq:hr_gamma}
\end{equation}
where $\alpha \in [0,1]$ is a learned scalar.

\noindent\textbf{Qualifier-conditioned propagation.}
The hidden state $\mathbf{h}_{uvqq'}^{(t)}$ encodes
all qualifier-conditioned paths from source $u$ to
node $v$ at layer $t$, where $q'$ denotes the query
relation and $q$ denotes the qualifier context
$\mathcal{Q}_q = \{(q_k^i, q_v^i)\}_{i=1}^n$
associated with the query. It is initialised at the
source node as:
\begin{equation}
  \mathbf{h}_{uvqq'}^{(0)} =
  \begin{cases}
    \mathbf{e}_{q'}^{\mathrm{qry}}
    + \mathbf{W}_{\mathrm{proj}}
    \sum_{(q_k,q_v)\in\mathcal{Q}_q}
    \mathbf{e}_{q_k}^{\mathrm{qry}}
    \odot \mathbf{e}_{q_v}^{\mathrm{qry}}
    & v = u, \\
    \mathbf{0} & \text{otherwise,}
  \end{cases}
  \label{eq:hr_init_inst}
\end{equation}


where $\mathbf{e}_{q'}^{\mathrm{qry}}$ is the
query relation embedding and
$\mathbf{e}_{q_k}^{\mathrm{qry}},
\mathbf{e}_{q_v}^{\mathrm{qry}}$ are dedicated
query-side qualifier embeddings.
$\mathbf{W}_{\mathrm{proj}} \in
\mathbb{R}^{d \times d}$ projects the qualifier
context into relation embedding space. This extends
NBFNet~\cite{zhu2021neural} at two levels: (i) the
\textsc{Indicator} boundary condition
(Eq.~\ref{eq:hr_init_inst}) is conditioned on both
the query relation $q'$ and the query-level qualifier
context $q$; and (ii) the aggregation at each layer
(Eq.~\ref{eq:hr_agg}) incorporates the edge-level
qualifier vector $\mathbf{h}_q^{xrv}$ via
$\gamma(\mathbf{h}_r, \mathbf{h}_q^{xrv})$,
injecting flow-level metadata into every
Bellman-Ford iteration.
\begin{align}
  \mathbf{a}_v^{(t)}
  &= \sum_{(x,r,v)\in\mathcal{E}(v)}
     \mathbf{h}_{uxqq'}^{(t-1)}
     + \gamma(\mathbf{h}_r,\mathbf{h}_q^{xrv}),
  \label{eq:hr_agg} \\
  \mathbf{h}_{uvqq'}^{(t)}
  &= \mathrm{Dropout}(\mathrm{ReLU}(\mathrm{LN}(
     \mathbf{W}^{(t)}[\mathbf{h}_{uvqq'}^{(t-1)}
     \|\mathbf{a}_v^{(t)}])))
     + \mathbf{h}_{uvqq'}^{(0)},
  \label{eq:hr_update}
\end{align}
where the residual to $\mathbf{h}_{uvqq'}^{(0)}$
preserves the query qualifier signal across all layers.

\noindent\textbf{Scoring and training.}
After $L$ layers, candidate tails are scored by:
\begin{equation}
  f_{\mathrm{HR}}(h,q',t,\mathcal{Q})
  = \mathrm{MLP}_{\mathrm{HR}}\!\left(
      [\mathbf{h}_{htqq'}^{(L)}
      \|\mathbf{e}_{q'}^{\mathrm{qry}}]
    \right),
  \label{eq:hr_score}
\end{equation}
trained with margin ranking loss.
Since each $(h,q')$ query requires one full
Bellman--Ford pass, training groups triples by
$(h,q')$ and samples $k_{\max}$ tails per group.
The complete procedure is given in
Algorithm~\ref{alg:hrnbfnet}.

\begin{algorithm}[t]
\caption{HR-NBFNet: Inference}
\label{alg:hrnbfnet}
\begin{algorithmic}[1]
\Require source node $h$,
  query relation $q'$,
  query qualifiers $q = \mathcal{Q}_q =
  \{(q_k^i, q_v^i)\}_{i=1}^n$,
  graph $\mathcal{G}^+$, \#layers $L$
\Ensure score vector $\mathbf{s} \in \mathbb{R}^N$
  where $\mathbf{s}[v] =
  f_{\mathrm{HR}}(h, q', v, \mathcal{Q}_q)$
  for all $v \in \mathcal{V}$

\Statex \hfill $\triangleright$ \textit{Boundary condition}
\State $\mathbf{H}^{(0)} \leftarrow \mathbf{0}^{N\times d}$
\State $\boldsymbol{\phi}_q \leftarrow
  \sum_{(q_k^i, q_v^i) \in q}
  \mathbf{e}_{q_k^i}^{\mathrm{qry}}
  \odot \mathbf{e}_{q_v^i}^{\mathrm{qry}}$
  \hfill $\triangleright$ \textit{encode $q$}
\State $\mathbf{H}^{(0)}[h] \leftarrow
  \mathbf{e}^{\mathrm{qry}}[q'] +
  \mathbf{W}_{\mathrm{proj}}\boldsymbol{\phi}_q$
  \hfill \eqref{eq:hr_init_inst}
  $\triangleright$ \textit{inject $q'$ and $q$}
\State $\mathbf{H}^0_{\mathrm{res}} \leftarrow
  \mathbf{H}^{(0)}.\mathrm{clone}()$

\Statex \hfill $\triangleright$ \textit{Bellman-Ford iteration}
\For{$t = 1$ \textbf{to} $L$}
  \State Compute $\mathbf{h}_q$ per edge via
    Eq.~\eqref{eq:hr_hq}
    \hfill $\triangleright$ \textit{edge qualifier}
  \State Compute $\boldsymbol{\gamma}$ via
    Eq.~\eqref{eq:hr_gamma}
    \hfill $\triangleright$ \textit{relation--qualifier merge}
  \State $\mathbf{A}^{(t)} \leftarrow
    \mathrm{scatter\_add}(
    \mathbf{H}^{(t-1)}[\mathbf{u}]
    + \boldsymbol{\gamma},\, \mathbf{v})$
    \hfill \eqref{eq:hr_agg}
  \State $\mathbf{H}^{(t)} \leftarrow
    \mathrm{Dropout}(\mathrm{ReLU}(\mathrm{LN}(
    \mathbf{W}^{(t)}[\mathbf{H}^{(t-1)}
    \|\mathbf{A}^{(t)}])))
    + \mathbf{H}^0_{\mathrm{res}}$
    \hfill \eqref{eq:hr_update}
\EndFor

\Statex \hfill $\triangleright$ \textit{Scoring}
\State \Return $\mathbf{s} \leftarrow
  \mathrm{MLP}_{\mathrm{HR}}(
  [\mathbf{H}^{(L)}\|
  \mathbf{e}^{\mathrm{qry}}[q']
  .\mathrm{expand}(N,d)])$
  \hfill \eqref{eq:hr_score}
\end{algorithmic}
\end{algorithm}

\noindent\textbf{Time complexity.}
$O(L\,|\mathcal{T}|\,Q_{\max}\,d
+ |\mathcal{E}|\,d^2)$ per query.
\subsubsection{Multi-Task Hyper-NBFNet}
\label{ssec:mthrnbfnet}

MT-HR-NBFNet augments HR-NBFNet with two auxiliary
tasks, relation and qualifier-value prediction,
alongside tail prediction. All three tasks share a
\emph{single} Bellman-Ford pass, so multi-task
supervision adds only three lightweight MLP heads
with no extra graph-traversal cost.

\noindent\textbf{Three prediction heads.}
Using $\mathbf{H}^{(L)}$ from
Eqs.~\ref{eq:hr_init_inst}--\ref{eq:hr_update}:
\begin{align}
  f_{\mathrm{tail}}(t)
  &= \mathrm{MLP}_t\!\left(
      [\mathbf{H}^{(L)}[t]
      \| \mathbf{e}_{q'}^{\mathrm{qry}}]
    \right),
  \label{eq:mthr_tail} \\
  \hat{\mathbf{y}}_{q'}
  &= \mathrm{MLP}_{q'}\!\left(
      \mathbf{H}^{(L)}[h]
    \right)
  \in \mathbb{R}^{|\mathcal{R}|},
  \label{eq:mthr_rel} \\
  \hat{\mathbf{y}}_{q_v}
  &= \mathrm{MLP}_{q_v}\!\left(
      \sigma\!\left(
        \mathbf{W}_g [
          \mathbf{H}^{(L)}[h]
          \|
          E_{q_k}^{\mathrm{head}}[q_k^*]
        ]
      \right)
      \odot \mathbf{H}^{(L)}[h]
    \right),
  \label{eq:mthr_qv}
\end{align}
where $E_{q_k}^{\mathrm{head}}$ is a dedicated
qualifier-key embedding table and $q_k^*$ is the
target qualifier key. The joint loss is:
\begin{equation}
  \mathcal{L}_{\mathrm{MT\text{-}HR}}
  = \lambda_t\,\mathcal{L}_t
  + \lambda_{q'}\,\mathcal{L}_{q'}
  + \lambda_{q_v}\,\mathcal{L}_{q_v},
  \label{eq:mthr_loss}
\end{equation}
where $\mathcal{L}_t$ is margin ranking, and
$\mathcal{L}_{q'}, \mathcal{L}_{q_v}$ are
cross-entropy. $\lambda_{q'} < \lambda_t$ since
propagated representations are already partially
discriminative for relation type.

\begin{algorithm}[t]
\caption{MT-HR-NBFNet: Training Step}
\label{alg:mthrnbfnet}
\begin{algorithmic}[1]
\Require group $(h, q')$, positives
  $\{t_i^+\}_{i=1}^{B}$,
  qualifiers $\{\mathcal{Q}_i\}$,
  graph $\mathcal{G}^+$,
  weights $\lambda_t, \lambda_{q'},
  \lambda_{q_v}$, margin $\delta$
\Ensure updated $\Theta$

\Statex \hfill $\triangleright$
  \textit{Single shared BF pass}
\State $\mathbf{H}^{(L)} \leftarrow
  \mathrm{HR\text{-}Infer}(h, q',
  \mathcal{Q}_{\mathrm{rep}},
  \mathcal{G}^+)$
  \hfill Alg.~\ref{alg:hrnbfnet}

\Statex \hfill $\triangleright$
  \textit{Task 1: tail (margin ranking)}
\State $\mathcal{L}_t \leftarrow
  \frac{1}{B}\sum_i \max(0,\, \delta
  - f_{\mathrm{tail}}(t_i^+)
  + f_{\mathrm{tail}}(t_i^-))$
  \hfill \eqref{eq:mthr_tail}

\Statex \hfill $\triangleright$
  \textit{Task 2: relation (cross-entropy)}
\State $\mathcal{L}_{q'} \leftarrow
  \mathrm{CE}(\mathrm{MLP}_{q'}(
  \mathbf{H}^{(L)}[h]),\, q')$
  \hfill \eqref{eq:mthr_rel}

\Statex \hfill $\triangleright$
  \textit{Task 3: qualifier-value (cross-entropy)}
\State $\mathcal{L}_{q_v} \leftarrow
  \mathrm{CE}(\hat{\mathbf{y}}_{q_v},\, q_v^*)$
  if $\mathcal{Q}_{\mathrm{rep}} \neq \emptyset$,
  else $0$
  \hfill \eqref{eq:mthr_qv}

\Statex \hfill $\triangleright$
  \textit{Combined loss and update}
\State $\Theta \leftarrow \mathrm{Adam}(
  \nabla_\Theta(\lambda_t \mathcal{L}_t
  + \lambda_{q'} \mathcal{L}_{q'}
  + \lambda_{q_v} \mathcal{L}_{q_v}),\,
  \mathrm{clip}{=}1.0)$
\State \Return $\Theta$
\end{algorithmic}
\end{algorithm}

\noindent\textbf{Time complexity.}
One step runs a single BF pass shared across all
three tasks:
$O(L\,|\mathcal{T}|\,Q_{\max}\,d
+ |\mathcal{E}|\,d^2)$,
asymptotically identical to HR-NBFNet.

\subsection{Paradigm~II: AlertStar}
\label{ssec:alertstar}
\begin{figure}[t]
    \centering
    \includegraphics[width=\linewidth]{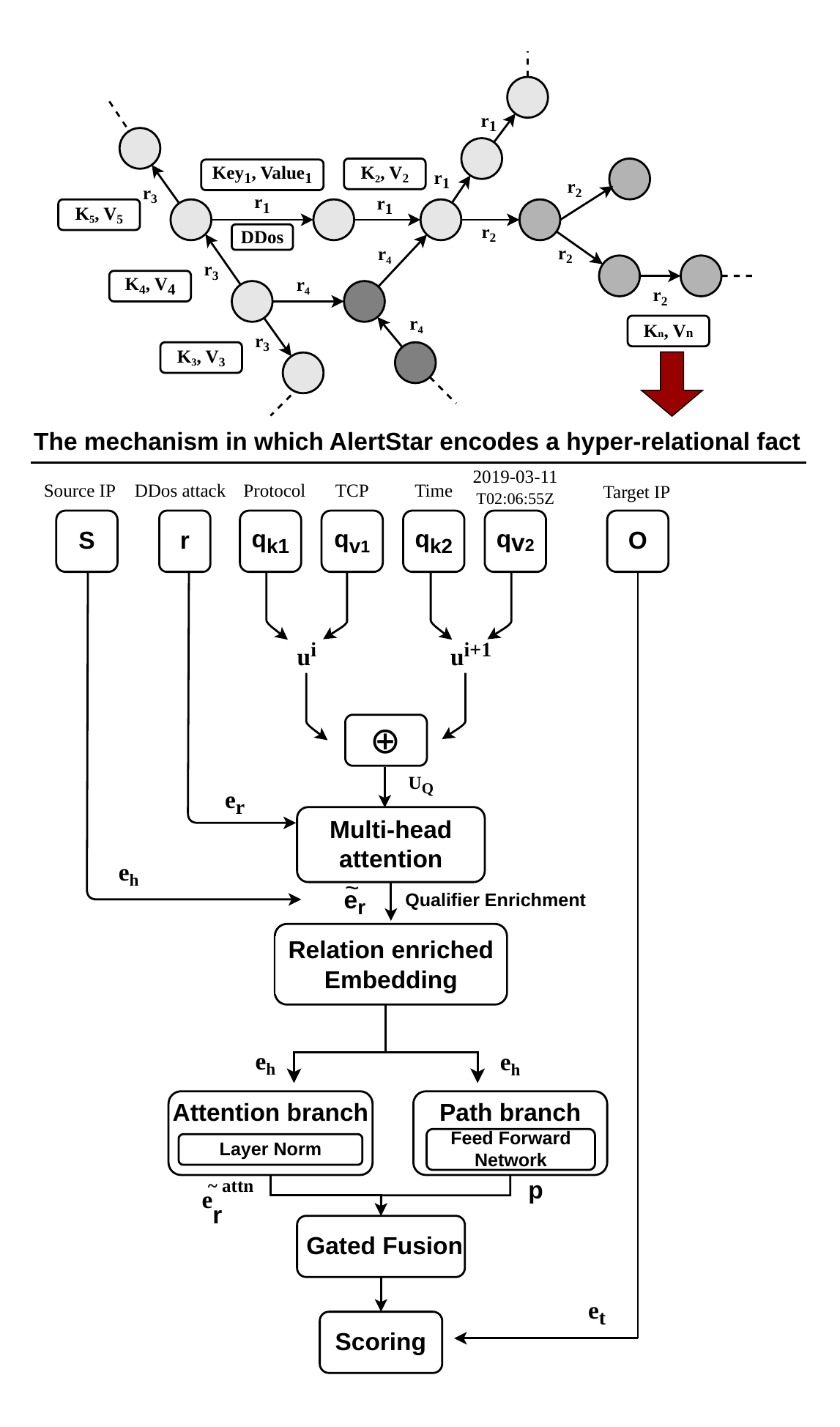}
    \caption{Architecture of AlertStar. Qualifier pairs
    are aggregated into $\mathbf{U}_\mathcal{Q}$ and
    used as key/value in MHA with $\mathbf{e}_r$ as
    query, producing $\tilde{\mathbf{e}}_r$. Two
    branches --- cross-attention and path-composition
    --- are fused via a trainable gate $\alpha =
    \sigma(g)$ into $\mathbf{z}$, scored against
    candidate tails via dot product.}
    \label{fig:alertstar}
\end{figure}

As illustrated in Figure \ref{fig:alertstar}, AlertStar fuses qualifier context and structural path
information entirely in embedding space for a single
triple $(h, r, \mathcal{Q})$, no graph traversal is
performed. It is orders of magnitude faster than
HR-NBFNet while retaining full qualifier awareness.

\noindent\textbf{Qualifier enrichment.}
Each qualifier pair is combined as
$\mathbf{u}^i = \mathbf{e}_{q_k^i} +
\mathbf{e}_{q_v^i}$, forming
$\mathbf{U}_\mathcal{Q} \in \mathbb{R}^{n \times d}$.
The relation is enriched via Multi-head cross-attention (MHA):
\begin{equation}
  \tilde{\mathbf{e}}_r =
  \begin{cases}
    \mathrm{MHA}(\mathbf{e}_r,
    \mathbf{U}_\mathcal{Q},
    \mathbf{U}_\mathcal{Q})
    & \mathcal{Q} \neq \emptyset, \\
    \mathbf{e}_r & \text{otherwise.}
  \end{cases}
  \label{eq:qual_enrich}
\end{equation}

\noindent\textbf{Cross-attention branch.}
Fuses $\mathbf{e}_h$ with $\tilde{\mathbf{e}}_r$ via
residual connection, followed by Layer
Normalisation~(LN), preserving the attacker IP
signal:
\begin{equation}
  \tilde{\mathbf{e}}_r^{\mathrm{attn}}
  = \mathrm{LN}(\mathbf{e}_h + \tilde{\mathbf{e}}_r).
  \label{eq:as_attn}
\end{equation}

\noindent\textbf{Path-composition branch.}
Models implicit compositional head--relation structure
without graph traversal:
\begin{equation}
  \mathbf{p}
  = \mathrm{LN}\!\left(
      \mathbf{e}_h
      + \mathrm{FFN}_{\mathrm{path}}\!\left(
          [\mathbf{e}_h
          \| \tilde{\mathbf{e}}_r^{\mathrm{attn}}]
        \right)
    \right),
  \label{eq:as_path}
\end{equation}
where $\mathrm{FFN}_{\mathrm{path}}$ is a
two-layer feed-forward network consisting of
$\mathrm{Linear}(2d{\to}d)$, Layer Normalisation,
ReLU activation, Dropout, and
$\mathrm{Linear}(d{\to}d)$.

\noindent\textbf{Gated fusion and scoring.}
A trainable scalar gate $g \in \mathbb{R}$
(initialised to $0.5$) balances both branches
via sigmoid $\sigma$:
\begin{equation}
  \alpha = \sigma(g), \quad
  \mathbf{z} = \alpha\,
  \tilde{\mathbf{e}}_r^{\mathrm{attn}}
  + (1{-}\alpha)\,\mathbf{p}
  \in \mathbb{R}^d,
  \label{eq:as_gate}
\end{equation}
where $\mathbf{z}$ is the fused alert representation.
Candidate tails are scored via dot product against
the entity embedding matrix
$E \in \mathbb{R}^{N \times d}$ ---
note that $t$ is not an input but a scoring target:
\begin{equation}
  f_{\mathrm{AS}}(h,r,t,\mathcal{Q})
  = \mathrm{Dropout}(\mathbf{z})^{\top}
  \mathbf{e}_t, \quad
  \mathbf{s} = \mathbf{z}\,E^{\top}
  \in \mathbb{R}^N,
  \label{eq:as_score}
\end{equation}
where $\mathbf{e}_t \in \mathbb{R}^d$ is the
embedding of candidate tail $t$ and
$N = |\mathcal{E}|$ is the number of entities.

\begin{algorithm}[t]
\caption{AlertStar: Forward Pass}
\label{alg:alertstar}
\begin{algorithmic}[1]
\Require head $h$, relation $r$,
  qualifiers $\mathcal{Q}$,
  optional tail $t^+$ (training only)
\Ensure score $f \in \mathbb{R}$ or
  $\mathbf{s} \in \mathbb{R}^N$

\Statex \hfill $\triangleright$
  \textit{Qualifier enrichment}
\State $\mathbf{U}_\mathcal{Q} \leftarrow
  [E_{q_k}[q_k^i] + E_{q_v}[q_v^i]]_{i=1}^n$
\State $\tilde{\mathbf{e}}_r \leftarrow
  \mathrm{MHA}(\mathbf{e}_r,
  \mathbf{U}_\mathcal{Q},
  \mathbf{U}_\mathcal{Q})$
  if $n>0$, else $\mathbf{e}_r$
  \hfill \eqref{eq:qual_enrich}

\Statex \hfill $\triangleright$
  \textit{Cross-attention branch}
\State $\tilde{\mathbf{e}}_r^{\mathrm{attn}}
  \leftarrow \mathrm{LN}(
  \mathbf{e}_h + \tilde{\mathbf{e}}_r)$
  \hfill \eqref{eq:as_attn}

\Statex \hfill $\triangleright$
  \textit{Path-composition branch}
\State $\mathbf{p} \leftarrow \mathrm{LN}(
  \mathbf{e}_h + \mathrm{FFN}_{\mathrm{path}}(
  [\mathbf{e}_h \|
  \tilde{\mathbf{e}}_r^{\mathrm{attn}}]))$
  \hfill \eqref{eq:as_path}

\Statex \hfill $\triangleright$
  \textit{Gated fusion}
\State $\mathbf{z} \leftarrow
  \sigma(g)\,\tilde{\mathbf{e}}_r^{\mathrm{attn}}
  + (1{-}\sigma(g))\,\mathbf{p}$
  \hfill \eqref{eq:as_gate}

\Statex \hfill $\triangleright$
  \textit{Scoring}
\If{training}
  \State \Return $\mathbf{z}^\top \mathbf{e}_{t^+}$
\Else
  \State \Return $\mathbf{z}\,E^\top$
\EndIf
\hfill \eqref{eq:as_score}
\end{algorithmic}
\end{algorithm}

\noindent\textbf{Time complexity.}
$O(Mnd + d^2 + Nd)$ per triple --- no graph traversal.
Typically $10^3$--$10^4{\times}$ faster per sample
than HR-NBFNet on dense alert graphs.


\subsubsection{MT-AlertStar: Multi-Task AlertStar}
\label{ssec:mtas}
\begin{figure}[t]
    \centering
    \includegraphics[width=\linewidth]{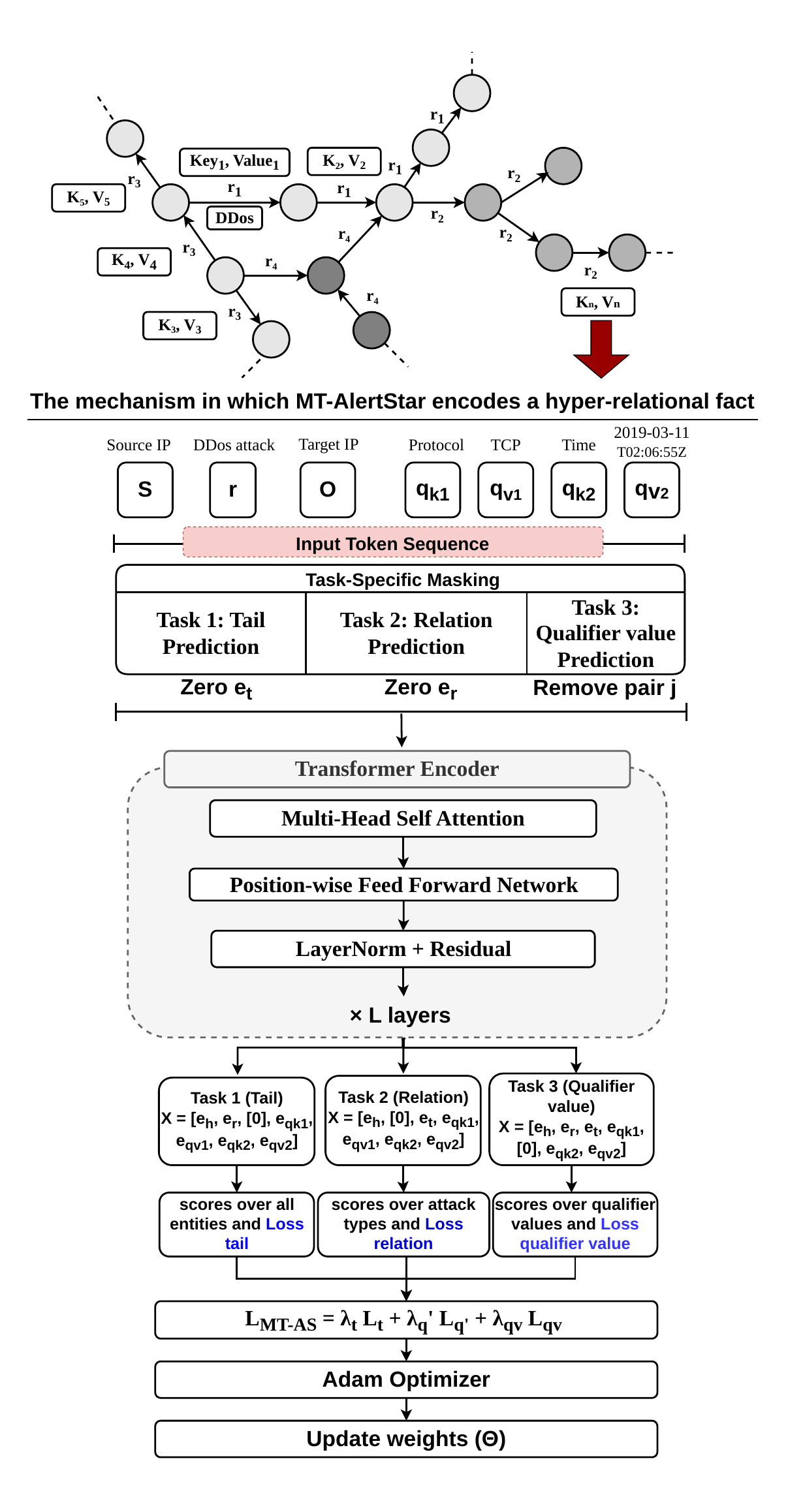}
    \caption{Architecture of MT-AlertStar. The masked
    token sequence $\mathbf{X}$ is encoded by a
    Transformer, whose relation-token output
    $\mathbf{h}_r = \mathbf{H}[1]$ serves as shared
    context for three MLP heads predicting tail,
    relation, and qualifier value jointly.}
    \label{fig:mtas}
\end{figure}

As illustrated in Figure \ref{fig:mtas}, MT-AlertStar extends AlertStar with a Transformer encoder \cite{varghese2024transformerg2g},
and three simultaneous prediction objectives, tail,
relation, and qualifier-value, mirroring MT-HR-NBFNet
within the embedding-based paradigm.

\noindent\textbf{Input and masking.}
For a qualified triple $(h, r, t, \mathcal{Q})$
with $n = |\mathcal{Q}|$ qualifier pairs, all
components are embedded into $\mathbb{R}^d$ and
arranged into a token sequence $\mathbf{X} \in \mathbb{R}^{(3+2n)\times d}$:
\begin{equation}
  \mathbf{X} = [
    \mathbf{e}_h,\, \mathbf{e}_r,\, \mathbf{e}_t,\,
    \mathbf{e}_{q_k^1},\, \mathbf{e}_{q_v^1},\,
    \ldots,\,
    \mathbf{e}_{q_k^n},\, \mathbf{e}_{q_v^n}]
  \in \mathbb{R}^{(3+2n)\times d},
  \label{eq:mt_input}
\end{equation}
where $\mathbf{e}_{q_k^i}, \mathbf{e}_{q_v^i}
\in \mathbb{R}^d$ are learned embeddings for
the $i$-th qualifier key and value respectively.
Task-specific masking prevents target leakage:
$\mathbf{e}_t \leftarrow \mathbf{0}$ for tail
prediction; $\mathbf{e}_r \leftarrow \mathbf{0}$
for relation prediction; qualifier pair $j$ is
removed for qualifier-value prediction.

\noindent\textbf{Transformer encoder.}
The masked sequence is passed through a standard
Transformer encoder \cite{varghese2024transformerg2g}:
$\mathbf{H} = \mathrm{TransformerEncoder}(\mathbf{X})
\in \mathbb{R}^{(3+2n) \times d}$,
where each layer applies multi-head self-attention,
position-wise FFN, LayerNorm, and residual
connections. The relation-token output
$\mathbf{h}_r = \mathbf{H}[1] \in \mathbb{R}^d$
--- position~1 in the sequence --- serves as
shared context for all three prediction heads,
as it attends over all other tokens including
the head entity, qualifier pairs, and masked
target position.

\noindent\textbf{Three prediction heads.}
All heads share the same architecture: two linear
layers with LayerNorm, ReLU, and Dropout
interleaved, mapping $\mathbb{R}^d \to
\mathbb{R}^{d_{\mathrm{out}}}$. Using the shared
context $\mathbf{h}_r$, each head produces a
score vector over its prediction space:
\begin{align}
  \hat{\mathbf{y}}_t &= \mathrm{MLP}_t(\mathbf{h}_r)
    \in \mathbb{R}^N,
  \label{eq:mt_tail} \\
  \hat{\mathbf{y}}_{q'} &= \mathrm{MLP}_{q'}(\mathbf{h}_r)
    \in \mathbb{R}^{|\mathcal{R}|},
  \label{eq:mt_rel} \\
  \hat{\mathbf{y}}_{q_v} &= \mathrm{MLP}_{q_v}(\mathbf{h}_r)
    \in \mathbb{R}^{|\mathcal{Q}_V|},
  \label{eq:mt_qval}
\end{align}
where $\hat{\mathbf{y}}_t$, $\hat{\mathbf{y}}_{q'}$,
and $\hat{\mathbf{y}}_{q_v}$ are the predicted score
vectors over $N$ entities, $|\mathcal{R}|$ attack
categories, and $|\mathcal{Q}_V|$ qualifier values
respectively. The three tasks are trained jointly:
\begin{equation}
  \mathcal{L}_{\mathrm{MT\text{-}AS}}
  = \lambda_t\,\mathcal{L}_t
  + \lambda_{q'}\,\mathcal{L}_{q'}
  + \lambda_{q_v}\,\mathcal{L}_{q_v},
  \label{eq:mt_loss}
\end{equation}
where $\mathcal{L}_t$, $\mathcal{L}_{q'}$, and
$\mathcal{L}_{q_v}$ are cross-entropy losses for
tail, relation, and qualifier-value prediction
respectively, and $\lambda_t, \lambda_{q'},
\lambda_{q_v} \in \mathbb{R}^+$ are task-specific
loss weights.

\begin{algorithm}[t]
\caption{MT-AlertStar: Training Step}
\label{alg:mtas}
\begin{algorithmic}[1]
\Require triple $(h, r, t, \mathcal{Q})$,
  task $\tau \in \{
  \texttt{tail}, \texttt{rel}, \texttt{qval}\}$
\Ensure updated $\Theta$

\Statex \hfill $\triangleright$
  \textit{Build masked token sequence}
\State $\mathbf{x}_h, \mathbf{x}_r, \mathbf{x}_t
  \leftarrow E[h], E_r[r], E[t]$
\State Mask: $\mathbf{x}_t{\leftarrow}\mathbf{0}$
  if \texttt{tail};\;
  $\mathbf{x}_r{\leftarrow}\mathbf{0}$
  if \texttt{rel};\;
  remove pair $j$ if \texttt{qval}
\State $\mathbf{X} \leftarrow
  [\mathbf{x}_h;\mathbf{x}_r;\mathbf{x}_t;\,
  \mathbf{e}_{q_k^1};\mathbf{e}_{q_v^1};
  \ldots]$
  \hfill \eqref{eq:mt_input}

\Statex \hfill $\triangleright$
  \textit{Transformer encoder}
\State $\mathbf{h}_r \leftarrow
  \mathrm{TransformerEncoder}(\mathbf{X})[1]$

\Statex \hfill $\triangleright$
  \textit{Task-specific head and loss}
\State $\hat{\mathbf{y}} \leftarrow
  \mathrm{MLP}_\tau(\mathbf{h}_r)$
  \hfill \eqref{eq:mt_tail}--\eqref{eq:mt_qval}
\State $\mathcal{L}_\tau \leftarrow
  \lambda_\tau \cdot
  \mathrm{CE}(\hat{\mathbf{y}},\, y^*)$

\Statex \hfill $\triangleright$
  \textit{Update}
\State $\Theta \leftarrow
  \mathrm{Adam}(\nabla_\Theta
  \mathcal{L}_\tau,\,\mathrm{clip}{=}1.0)$
\State \Return $\Theta$
\end{algorithmic}
\end{algorithm}

\noindent\textbf{Time complexity.}
$O(n^2 d + ndF + Nd)$ per sample, no graph
traversal.

\subsection{Paradigm~III: Complex Hyper-Relational}
\label{ssec:nbfstarqe}
\begin{figure*}[t]
    \centering
    \includegraphics[width=\linewidth]{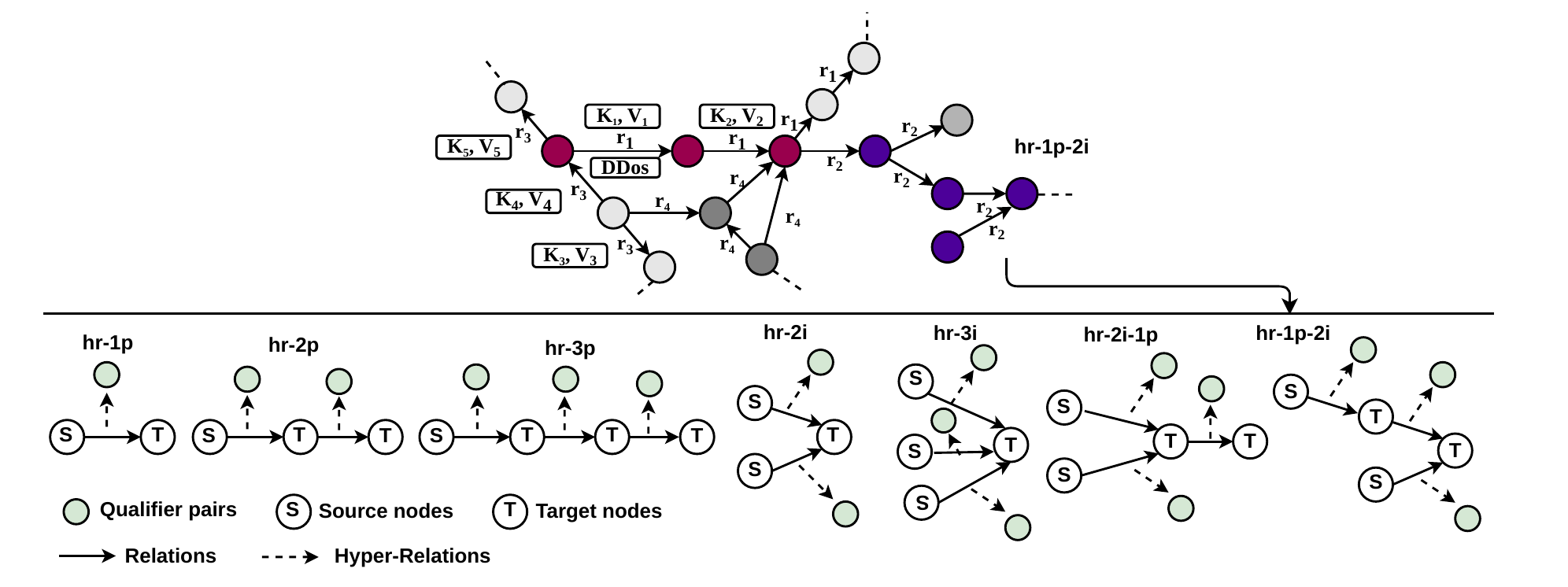}
    \caption{Hyper-relational query templates.
    Qualifier pairs on each edge may vary from $0$ to
    $n$.}
    \label{fig:CQ}
\end{figure*}

As illustrated in Figure \ref{fig:CQ}, HR-NBFNet-CQ answers complex first-order logic queries
over the hyper-relational alert graph, enabling richer
threat intelligence beyond single-hop link prediction.

\noindent\textbf{Query types.}
Four templates are supported, each grounded in a
concrete threat scenario:

\noindent\textbf{1p} $(?t.\; r(h, t, \mathcal{Q}_1))$
--- direct prediction. \textit{``Which IP was targeted
by 185.192.59.136 via DDoS with flow count 17,094 over
TCP on port 22?''}

\noindent\textbf{2p} $(?t.\; \exists y:\;
r_1(h,y,\mathcal{Q}_1) \wedge r_2(y,t,\mathcal{Q}_2))$
--- lateral movement: attacker first compromises an
intermediate host $y$, which then attacks $t$.
$y$ is inferred implicitly during propagation.
\textit{``185.192.59.136 performed a Recon Scan on
$y$; $y$ then launched a DoS --- who was the final
target?''}

\noindent\textbf{2i} $(?t.\;
r_1(h_1,t,\mathcal{Q}_1) \wedge
r_2(h_2,t,\mathcal{Q}_2))$
--- coordinated attack: finds a shared victim
targeted by two source IPs simultaneously.
\textit{``Which IP was hit by both 185.192.59.136
via DDoS and 78.234.46.141 via Recon Scan?''}

\noindent\textbf{2u} $(?t.\;
r_1(h_1,t,\mathcal{Q}_1) \vee
r_2(h_2,t,\mathcal{Q}_2))$
--- campaign victim set: retrieves all IPs targeted
by either attacker. Unlike StarQE~\cite{alivanistos2021query},
which restricts to conjunctive templates, we support
this disjunctive query motivated by campaign-level
co-targeting in Warden, where coordinated attacks
against overlapping victim sets are operationally
meaningful for CERT analysts.

\noindent\textbf{Composition operator.}
All query types share the residual path-composition
operator $\phi_{\mathrm{NBF}}$, which models one
hop entirely in embedding space. Given a current
node embedding $\mathbf{x} \in \mathbb{R}^d$,
a relation $r$, qualifier set $\mathcal{Q}$, and
an anchor embedding $\mathbf{x}_0 \in \mathbb{R}^d$
(the source entity, preserved via residual):
\begin{equation}
  \phi_{\mathrm{NBF}}(\mathbf{x}, r,
  \mathcal{Q}, \mathbf{x}_0)
  = \mathbf{x}_0
    + \mathrm{FFN}_\phi([\mathbf{x}
    \| \tilde{\mathbf{e}}_r]),
  \label{eq:nbfqe_compose}
\end{equation}
where $[\cdot\|\cdot]$ denotes concatenation,
$\tilde{\mathbf{e}}_r$ is the qualifier-enriched
relation embedding from Eq.~\ref{eq:qual_enrich},
and $\mathrm{FFN}_\phi{:}\,\mathbb{R}^{2d}
\to \mathbb{R}^d$ comprises two linear layers
with LayerNorm, ReLU, and Dropout.

\noindent\textbf{Query formulations.}
Each query type produces a query vector
$\mathbf{q} \in \mathbb{R}^d$ by composing
$\phi_{\mathrm{NBF}}$ in different structural
patterns. For 2p, $r_1$ and $r_2$ are two
sequential relations and $\emptyset$ denotes
an empty qualifier set on the second hop.
For 2i and 2u, $h_1$ and $h_2$ are two anchor
source IPs with relations $r_1$ and $r_2$:
\begin{align}
  \mathbf{q}^{1p}
  &= \phi_{\mathrm{NBF}}(\mathbf{e}_h, r,
    \mathcal{Q}, \mathbf{e}_h),
  \label{eq:nbfqe_1p} \\
  \mathbf{q}^{2p}
  &= \phi_{\mathrm{NBF}}(
    \phi_{\mathrm{NBF}}(\mathbf{e}_h, r_1,
    \mathcal{Q}, \mathbf{e}_h),
    r_2, \emptyset, \mathbf{e}_h),
  \label{eq:nbfqe_2p} \\
  \mathbf{q}^{2i}
  &= W_i[\phi_{\mathrm{NBF}}(\mathbf{e}_{h_1},
    r_1, \mathcal{Q}, \mathbf{e}_{h_1})
    \| \phi_{\mathrm{NBF}}(\mathbf{e}_{h_2},
    r_2, \emptyset, \mathbf{e}_{h_2})],
  \label{eq:nbfqe_2i} \\
  \mathbf{q}^{2u}
  &= \tfrac{1}{2}(
    \phi_{\mathrm{NBF}}(\mathbf{e}_{h_1},
    r_1, \mathcal{Q}, \mathbf{e}_{h_1})
    + \phi_{\mathrm{NBF}}(\mathbf{e}_{h_2},
    r_2, \emptyset, \mathbf{e}_{h_2})),
  \label{eq:nbfqe_2u}
\end{align}
where $W_i \in \mathbb{R}^{d \times 2d}$ is a
learned projection for asymmetric intersection.
All query types are scored by dot product against
candidate entity embedding
$\mathbf{e}_e \in \mathbb{R}^d$:
$f = \mathrm{Dropout}(\mathbf{q})^\top \mathbf{e}_e$,
with inference $\mathbf{s} = \mathbf{q} E^\top
\in \mathbb{R}^N$ over all $N$ entities costing
$O(Nd)$.

\begin{algorithm}[t]
\caption{HR-NBFNet-CQ: Training Step}
\label{alg:nbfstarqe}
\begin{algorithmic}[1]
\Require triple $(h, r, t^+, \mathcal{Q})$,
  neighbour index $\mathcal{I}_{\mathrm{out}}$,
  anchor index $\mathcal{I}_{\mathrm{in}}$,
  margin $\delta$
\Ensure updated $\Theta$

\Statex \hfill $\triangleright$
  \textit{Qualifier enrichment}
\State $\tilde{\mathbf{e}}_r \leftarrow
  \mathrm{QualEnrich}(E_r[r], \mathcal{Q})$
  \hfill \eqref{eq:qual_enrich}
\State $t^- \leftarrow \mathrm{Uniform}(\mathcal{E})$;\;
  $\mathcal{A} \leftarrow \emptyset$;\;
  $\mathcal{L} \leftarrow 0$

\Statex \hfill $\triangleright$
  \textit{1p (always)}
\State $\mathbf{q}^{1p} \leftarrow
  \phi_{\mathrm{NBF}}(E[h], r,
  \mathcal{Q}, E[h])$
  \hfill \eqref{eq:nbfqe_1p}
\State $\mathcal{L} \mathrel{+}=
  \max(0, \delta - \mathbf{q}^{1p\top}E[t^+]
  + \mathbf{q}^{1p\top}E[t^-])$;\;
  $\mathcal{A} \mathrel{+}= \{1p\}$

\Statex \hfill $\triangleright$
  \textit{2p (if chain exists)}
\If{$\mathcal{I}_{\mathrm{out}}[t^+] \neq \emptyset$}
  \State Sample $(r_2, t_2)$;\;
  $\mathbf{q}^{2p} \leftarrow
  \phi_{\mathrm{NBF}}(\mathbf{q}^{1p},
  r_2, \emptyset, E[h])$
  \hfill \eqref{eq:nbfqe_2p}
  \State $\mathcal{L} \mathrel{+}=
    \max(0, \delta -
    \mathbf{q}^{2p\top}E[t_2] +
    \mathbf{q}^{2p\top}E[t^-])$;\;
    $\mathcal{A} \mathrel{+}= \{2p\}$
\EndIf

\Statex \hfill $\triangleright$
  \textit{2i (if second anchor exists)}
\If{$\exists (h_2,r_2) \in
  \mathcal{I}_{\mathrm{in}}[t^+],\, h_2 \neq h$}
  \State $\mathbf{q}^{2i} \leftarrow
    W_i[\mathbf{q}^{1p} \|
    \phi_{\mathrm{NBF}}(E[h_2],
    r_2, \emptyset, E[h_2])]$
    \hfill \eqref{eq:nbfqe_2i}
  \State $\mathcal{L} \mathrel{+}=
    \max(0, \delta -
    \mathbf{q}^{2i\top}E[t^+] +
    \mathbf{q}^{2i\top}E[t^-])$;\;
    $\mathcal{A} \mathrel{+}= \{2i\}$
\EndIf

\Statex \hfill $\triangleright$
  \textit{Average and update}
\State $\Theta \leftarrow \mathrm{Adam}(
  \nabla_\Theta(\mathcal{L}/|\mathcal{A}|),\,
  \mathrm{clip}{=}1.0)$
\State \Return $\Theta$
\end{algorithmic}
\end{algorithm}

\noindent\textbf{Time complexity.}
$O(nd + d^2 + Nd)$ per triple --- same class as
AlertStar, much cheaper than HR-NBFNet.

\subsection{Shared Training Objective}
\label{ssec:training}

All five models are trained with Adam, gradient
clipping to unit norm, and model selection by best
validation MRR. The shared margin ranking loss is:
\begin{equation}
  \mathcal{L}_{\mathrm{margin}}
  = \sum_{(h,r,t,\mathcal{Q})}
    \max\!\left(0,\;
      \delta
      - f(h,r,t,\mathcal{Q})
      + f(h,r,t^-,\mathcal{Q})
    \right),
  \label{eq:margin_loss_shared}
\end{equation}
where $t^- \sim \mathrm{Uniform}(\mathcal{E})$.
MT-HR-NBFNet and MT-AlertStar use the joint objectives
of Eqs.~\ref{eq:mthr_loss} and~\ref{eq:mt_loss}.
HR-NBFNet-CQ averages the margin loss over active
query types. Table~\ref{tab:complexity} summarises
per-sample inference complexity.

\begin{table}[t]

\centering
\caption{Per-sample inference complexity.
  $N{=}|\mathcal{E}|$, $E{=}|\mathcal{T}|$,
  $L{=}$layers, $Q{=}Q_{\max}$,
  $n{=}|\mathcal{Q}|$, $S{=}3{+}2n$,
  $F{=}$FFN width.}
\label{tab:complexity}
\renewcommand{\arraystretch}{1.2}
\begin{tabular}{lcc}

\toprule
\textbf{Model} & \textbf{Complexity}
  & \textbf{Graph Traversal?} \\
\midrule
AlertStar       & $O(nd + d^2 + Nd)$        & No  \\
MT-AlertStar    & $O(S^2d + SdF + Nd)$      & No  \\
HR-NBFNet-CQ    & $O(nd + d^2 + Nd)$        & No  \\
HR-NBFNet       & $O(LEQd + Nd^2)$          & Yes \\
MT-HR-NBFNet    & $O(LEQd + Nd^2)$          & Yes \\
\bottomrule
\end{tabular}
\end{table}


\begin{table*}[t]
\tiny
\centering
\caption{KGC results on the \textbf{Warden Alert dataset} 
         using \textbf{statements} under \textbf{inductive} and \textbf{transductive} settings.}
\label{tab:paradiam_compare}
\begin{tabular}{l ccccc ccccc}
\toprule
 & \multicolumn{5}{c}{\textbf{Inductive}} & \multicolumn{5}{c}{\textbf{Transductive}} \\
\cmidrule(lr){2-6} \cmidrule(lr){7-11}
\textbf{Model} & \textbf{MR} $\downarrow$ & \textbf{MRR} $\uparrow$ & \textbf{H@1} $\uparrow$ & \textbf{H@3} $\uparrow$& \textbf{H@10} $\uparrow$ & \textbf{MR} $\downarrow$ & \textbf{MRR} $\uparrow$& \textbf{H@1} $\uparrow$ & \textbf{H@3} $\uparrow$ & \textbf{H@10} $\uparrow$ \\
\midrule
\multicolumn{11}{c}{\textit{\textbf{Paradigm I: Graph Propagation}}} \\
\cmidrule{1-11}

HR-NBFNet                   & 4616.2 & 0.4666 & 0.4332 & 0.4791 & 0.5283 & 4182.4 & 0.4785 & 0.4408 & 0.4956 & 0.5465 \\
MultiTask HR-NBFNet         & 5229.5 & 0.4729 & 0.4373 & 0.4849 & 0.5528 & 589.8 & 0.4134 & 0.3782 & 0.4171 & 0.4926 \\
\midrule

\midrule
\multicolumn{11}{c}{\textit{\textbf{Paradigm II: Embedding-Based}}} \\
\midrule
AlertStar                        & 2645.2 & 0.4280 & 0.3400 & \textbf{0.4900} & \textbf{0.5620} & 3083.1 & 0.4620 & 0.3980 & \textbf{0.5040} & \textbf{0.5640}\\
MultiTask AlertStar         & \textbf{3016.4} & \textbf{0.5618} & \textbf{0.5160} & \textbf{0.5860} & \textbf{0.6460} & \textbf{3341.5} & \textbf{0.5384} & \textbf{0.5040} & \textbf{0.5480} & \textbf{0.6140}\\

\bottomrule
\end{tabular}
\end{table*}


\begin{table*}[t]
\tiny
\centering
\caption{KGC results on the \textbf{Warden Alert dataset} using \textbf{statements} under the \textbf{inductive} and \textbf{transductive} settings.}
\label{tab:kgc_statements_warden}
\begin{tabular}{l ccccc ccccc}
\toprule
 & \multicolumn{5}{c}{\textbf{Inductive}} & \multicolumn{5}{c}{\textbf{Transductive}} \\
\cmidrule(lr){2-6} \cmidrule(lr){7-11}
\textbf{Model} & \textbf{MR} $\downarrow$ & \textbf{MRR} $\uparrow$ & \textbf{H@1} $\uparrow$ & \textbf{H@3} $\uparrow$ & \textbf{H@10} $\uparrow$ & \textbf{MR} $\downarrow$ & \textbf{MRR} $\uparrow$ & \textbf{H@1} $\uparrow$ & \textbf{H@3} $\uparrow$ & \textbf{H@10} $\uparrow$ \\
\midrule

\multicolumn{11}{c}{\textit{33\%}} \\
\cmidrule{1-11}
ShrinkE         & 2724.0 & 0.4665 & 0.4380 & 0.4760 & 0.5220 & 3087.2 & 0.5012 & 0.4820 & 0.5060 & 0.5400 \\
StarE           & 2858.4 & 0.4205 & 0.3640 & 0.4520 & 0.5300 & 3262.7 & 0.4451 & 0.3940 & 0.4680 & 0.4940 \\
\textbf{AlertStar}       & 2765.3 & 0.4371 & 0.3520 & 0.\textbf{5040} & \textbf{0.5840} & 3031.6 & 0.4627 & 0.3840 & \textbf{0.5280} & \textbf{0.5920} \\
\textbf{MultiTask AlertStar} & \textbf{2680.2} & \textbf{0.5860} & \textbf{0.5400} & \textbf{0.6100} & \textbf{0.6850} & \textbf{3174.2} & \textbf{0.5582} & \textbf{0.5200} & \textbf{0.5750} & \textbf{0.6350} \\
\midrule

\multicolumn{11}{c}{\textit{66\%}} \\
\cmidrule{1-11}
ShrinkE         & 2920.0 & 0.4683 & 0.4440 & 0.4760 & 0.5100 & 2692.4 & 0.5026 & 0.4860 & 0.5040 & 0.5360 \\
StarE           & 3187.6 & 0.3209 & 0.2100 & 0.4080 & 0.5140 & 3098.7 & 0.3799 & 0.2720 & 0.4540 & 0.5340 \\
\textbf{AlertStar}       & 3111.1 & 0.4153 & 0.3420 & 0.4600 & \textbf{0.5420} & 3088.3 & 0.4666 & 0.3920 & \textbf{0.5260} & \textbf{0.5900} \\
\textbf{MultiTask AlertStar} & \textbf{2532.5} & \textbf{0.5968} & \textbf{0.5600} & \textbf{0.6100} & \textbf{0.6850} & \textbf{3329.0} & \textbf{0.5328} & \textbf{0.5000} & \textbf{0.5550} & \textbf{0.5800} \\
\midrule

\multicolumn{11}{c}{\textit{100\%}} \\
\cmidrule{1-11}
ShrinkE         & 2784.8 & 0.4567 & 0.4380 & 0.4460 & 0.5220 & 2701.1 & 0.5104 & 0.4900 & 0.5180 & 0.5540 \\
StarE           & 3522.9 & 0.4191 & 0.3360 & 0.4740 & 0.5480 & 3744.6 & 0.4227 & 0.3420 & 0.4840 & 0.5560 \\
\textbf{AlertStar}       & 3289.9 & 0.4254 & 0.3260 & \textbf{0.5060} & \textbf{0.5860} & 3180.5 & 0.4797 & 0.4160 & \textbf{0.5220} & \textbf{0.5860} \\
\textbf{MultiTask AlertStar} & \textbf{2977.9} & \textbf{0.5838} & \textbf{0.5400} & \textbf{0.6050} & \textbf{0.6750} & \textbf{3307.0} & \textbf{0.5477} & \textbf{0.5150} & \textbf{0.5650} & \textbf{0.6100} \\
\bottomrule
\end{tabular}
\end{table*}


\begin{table*}[t]
\tiny
\centering
\caption{KGC results on the \textbf{UNSW dataset} using \textbf{statements} under the \textbf{inductive} and \textbf{transductive} settings.}
\label{tab:kgc_statements_unsw}
\begin{tabular}{l ccccc ccccc}
\toprule
 & \multicolumn{5}{c}{\textbf{Inductive}} & \multicolumn{5}{c}{\textbf{Transductive}} \\
\cmidrule(lr){2-6} \cmidrule(lr){7-11}
\textbf{Model} & \textbf{MR} $\downarrow$ & \textbf{MRR} $\uparrow$ & \textbf{H@1} $\uparrow$ & \textbf{H@3} $\uparrow$ & \textbf{H@10} $\uparrow$ & \textbf{MR} $\downarrow$ & \textbf{MRR} $\uparrow$ & \textbf{H@1} $\uparrow$ & \textbf{H@3} $\uparrow$ & \textbf{H@10} $\uparrow$ \\
\midrule

\multicolumn{11}{c}{\textit{33\%}} \\
\cmidrule{1-11}
ShrinkE         & 4.2680 & 0.4220 & 0.2200 & 0.5000 & 1.000 & 4.2880 & 0.4083 & 0.2020 & 0.4800 & 1.000 \\
StarE           & 4.1460 & 0.4367 & 0.2280 & 0.5480 & 0.996 & 4.1560 & 0.4267 & 0.2160 & 0.5080 & 1.000 \\
\textbf{AlertStar}       & \textbf{3.9300} & \textbf{0.4497} & \textbf{0.2240} & \textbf{0.5640} & \textbf{1.000} & \textbf{4.0140} & \textbf{0.4394} & \textbf{0.2260} & \textbf{0.5300} & \textbf{1.000} \\
\textbf{MultiTask AlertStar} & \textbf{3.9350} & \textbf{0.4490} & \textbf{0.2350} & \textbf{0.5400} & \textbf{1.000} & \textbf{3.9250} & \textbf{0.4626} & \textbf{0.2600} & \textbf{0.5500} & \textbf{1.000} \\
\midrule

\multicolumn{11}{c}{\textit{66\%}} \\
\cmidrule{1-11}
ShrinkE         & 4.1020 & 0.4483 & 0.2440 & 0.5500 & 1.000 & 4.1500 & 0.4241 & 0.2100 & 0.5100 & 1.000 \\
StarE           & 4.2980 & 0.4296 & 0.2280 & 0.5280 & 0.996 & 4.0800 & 0.4404 & 0.2360 & 0.5060 & 1.000 \\
\textbf{AlertStar}       & \textbf{4.0160} & \textbf{0.4575} & \textbf{0.2440} & \textbf{0.5740} & \textbf{1.000} & \textbf{4.0160} & \textbf{0.4575} & \textbf{0.2440} & \textbf{0.5740} & \textbf{1.000} \\
\textbf{MultiTask AlertStar} & \textbf{3.6800} & \textbf{0.5023} & \textbf{0.3200} & \textbf{0.5800} & \textbf{1.000} & \textbf{3.7900} & \textbf{0.4752} & \textbf{0.2750} & \textbf{0.5750} & \textbf{1.000} \\
\midrule

\multicolumn{11}{c}{\textit{100\%}} \\
\cmidrule{1-11}
ShrinkE         & 4.0120 & 0.4449 & 0.2360 & 0.5680 & 1.000 & 4.0680 & 0.4477 & 0.2440 & 0.5500 & 1.000 \\
StarE           & 4.0020 & 0.4595 & 0.2460 & 0.5760 & 0.996 & 4.0460 & 0.4236 & 0.2040 & 0.5320 & 1.000 \\
\textbf{AlertStar}       & \textbf{3.8540} & \textbf{0.4707} & \textbf{0.2620} & \textbf{0.5900} & \textbf{1.000} & \textbf{3.8440} & \textbf{0.4647} & \textbf{0.2520} & \textbf{0.5820} & \textbf{1.000} \\
\textbf{MultiTask AlertStar} & \textbf{3.6450} & \textbf{0.4809} & \textbf{0.2700} & \textbf{0.6050} & \textbf{1.000} & \textbf{4.0000} & \textbf{0.4641} & \textbf{0.2650} & \textbf{0.5550} & \textbf{1.000} \\
\bottomrule
\end{tabular}
\end{table*}

\begin{table*}[t]
\centering
\tiny
\caption{Ablation A4: Model performance across qualifier density levels (inductive).}
\label{tab:ablation_density}
\begin{tabular}{l|ccc|ccc|ccc}
\toprule
 & \multicolumn{3}{c|}{\textbf{Q33\%}} 
 & \multicolumn{3}{c|}{\textbf{Q66\%}} 
 & \multicolumn{3}{c}{\textbf{Q100\%}} \\
\textbf{Model} 
 & \textbf{MRR} & \textbf{H@1} & \textbf{H@10}
 & \textbf{MRR} & \textbf{H@1} & \textbf{H@10}
 & \textbf{MRR} & \textbf{H@1} & \textbf{H@10} \\
\midrule
StarE               & 0.2783 & 0.0940 & 0.5140 
                    & 0.3442 & 0.2180 & 0.5140 
                    & 0.4051 & 0.3140 & 0.5340 \\
AlertStar           & 0.4282 & 0.3380 & 0.5760 
                    & 0.4167 & 0.3340 & 0.5440 
                    & 0.4313 & 0.3420 & 0.5820 \\
HyNT                & 0.5632 & 0.5140 & 0.6460 
                    & 0.5726 & 0.5320 & 0.6440 
                    & 0.5692 & 0.5280 & 0.6460 \\
MultiTask AlertStar & 0.5546 & 0.5040 & 0.6520 
                    & 0.5663 & 0.5200 & 0.6540 
                    & 0.5531 & 0.5060 & 0.6460 \\
\bottomrule
\end{tabular}
\end{table*}


\begin{table*}[t]
\tiny
\centering
\caption{KGC results on the \textbf{Warden Alert dataset} 
         using \textbf{triples} under \textbf{inductive} and \textbf{transductive} settings.}
\label{tab:kgc_triples_warden}
\begin{tabular}{l ccccc ccccc}
\toprule
 & \multicolumn{5}{c}{\textbf{Inductive}} & \multicolumn{5}{c}{\textbf{Transductive}} \\
\cmidrule(lr){2-6} \cmidrule(lr){7-11}
\textbf{Model} & \textbf{MR} $\downarrow$ & \textbf{MRR} $\uparrow$& \textbf{H@1} $\uparrow$& \textbf{H@3} $\uparrow$& \textbf{H@10} $\uparrow$& \textbf{MR} $\downarrow$& \textbf{MRR} $\uparrow$& \textbf{H@1} $\uparrow$& \textbf{H@3} $\uparrow$& \textbf{H@10} $\uparrow$\\
\midrule
ShrinkE         & 2699.4 & 0.4580 & 0.4140 & 0.4840 & 0.5320 & 2889.4 & 0.4998 & 0.4680 & 0.5180 & 0.5500 \\
StarE           & 2760.9 & 0.4098 & 0.3400 & 0.4680 & 0.5300 & 3150.9 & 0.4503 & 0.4000 & 0.4860 & 0.5460 \\
NBFNet          & 3616.5 & 0.4853 & 0.4481 & 0.5060 & 0.5480 & 4524.8 & 0.4818 & 0.4451 & 0.5019 & 0.5511 \\
\midrule
\textbf{AlertStar}       & \textbf{2561.0} & 0.4565 & 0.3720 & \textbf{0.5140} & \textbf{0.5920} & 2746.2 & 0.4862 & 0.4320 & \textbf{0.5140} & \textbf{0.5840} \\
\textbf{MultiTask AlertStar (Tail+Rel)} & \textbf{2430.0} & \textbf{0.5860} & \textbf{0.5450} & \textbf{0.6000} & \textbf{0.6750} & \textbf{3138.6} & \textbf{0.5762} & \textbf{0.5450} & \textbf{0.5900} & \textbf{0.6400} \\

\bottomrule
\end{tabular}
\end{table*}

\begin{table*}[t]
\tiny
\centering
\caption{KGC results on the \textbf{UNSW dataset} 
         using \textbf{triples} under \textbf{inductive} and \textbf{transductive} settings.}
\label{tab:kgc_triples_UNSW}
\begin{tabular}{l ccccc ccccc}
\toprule
 & \multicolumn{5}{c}{\textbf{Inductive}} & \multicolumn{5}{c}{\textbf{Transductive}} \\
\cmidrule(lr){2-6} \cmidrule(lr){7-11}
\textbf{Model} & \textbf{MR} $\downarrow$ & \textbf{MRR} $\uparrow$& \textbf{H@1} $\uparrow$& \textbf{H@3} $\uparrow$& \textbf{H@10} $\uparrow$& \textbf{MR} $\downarrow$& \textbf{MRR} $\uparrow$& \textbf{H@1} $\uparrow$& \textbf{H@3} $\uparrow$& \textbf{H@10} $\uparrow$\\
\midrule
ShrinkE         & 4.1380 & 0.4405 & 0.2400 & 0.5320 & 1.000 & 4.1360 & 0.4305 & 0.2220 & 0.5140 & 1.000 \\
StarE           & 4.1300 & 0.4445 & 0.2460 & 0.5240 & 0.996 & 4.1300 & 0.4163 & 0.1980 & 0.5120 & 1.000 \\
NBFNet          & 4.2405 & 0.4288 & 0.2252 & 0.5120 & 1.000 & 4.2484 & 0.4279 & 0.2228 & 0.5142 & 1.000 \\
\midrule
\textbf{AlertStar}       & \textbf{4.0060} & \textbf{0.4469} & 0.2320 & \textbf{0.5480} & \textbf{1.000} & 4.0060 & 0.4355 & 0.2220 & \textbf{0.5320} & \textbf{1.000} \\
\textbf{MultiTask AlertStar (Tail+Rel)} & \textbf{3.9500} & \textbf{0.4409} & \textbf{0.2150} & \textbf{0.5850} & \textbf{1.000} & \textbf{3.9700} & \textbf{0.4519} & \textbf{0.2550} & \textbf{0.5200} & \textbf{1.000} \\

\bottomrule
\end{tabular}
\end{table*}

\begin{table*}[t]
\tiny
\centering
\caption{Complex query answering results on the \textbf{Warden Alert dataset} under the \textbf{inductive} and \textbf{transductive} settings.}
\label{tab:complex_query_answering_warden}
\footnotesize
\resizebox{\textwidth}{!}{%
\begin{tabular}{l ccccc ccccc ccccc}
\toprule
 & \multicolumn{15}{c}{\textbf{Inductive}} \\ \toprule
 & \multicolumn{5}{c}{\textbf{33\%}} & \multicolumn{5}{c}{\textbf{66\%}} & \multicolumn{5}{c}{\textbf{100\%}} \\
\cmidrule(lr){2-6} \cmidrule(lr){7-11} \cmidrule(lr){12-16}
\textbf{Query}
  & \textbf{MR} $\downarrow$ & \textbf{MRR} $\uparrow$& \textbf{H@1} $\uparrow$& \textbf{H@3} $\uparrow$& \textbf{H@10} $\uparrow$
  & \textbf{MR} $\downarrow$& \textbf{MRR} $\uparrow$& \textbf{H@1} $\uparrow$& \textbf{H@3} $\uparrow$& \textbf{H@10} $\uparrow$
  & \textbf{MR} $\downarrow$& \textbf{MRR} $\uparrow$& \textbf{H@1} $\uparrow$& \textbf{H@3} $\uparrow$& \textbf{H@10} $\uparrow$\\
\midrule
\multicolumn{16}{c}{\textit{StarQE}} \\
\midrule
1p & 43.810 & 0.4875 & 0.460 & 0.475 & 0.555 &  79.305 & 0.4600 & 0.440 & 0.445 & 0.495 & 53.820 & 0.5103 & 0.4950 & 0.5050 & 0.5350 \\
2p & 100.66 & 0.0176 & 0.000 & 0.000 & 0.000 &  51.250 & 0.2611 & 0.250 & 0.250 & 0.250 & 34.545 & 0.6694 & 0.6363 & 0.7272 & 0.7272 \\
\rowcolor{gray!15}
2i & 26.545 & 0.3584 & 0.005 & 0.685 & 0.715 &  31.965 & 0.3454 & 0.010 & 0.655 & 0.700 & 33.575 & 0.6835 & 0.6750 & 0.6750 & 0.6900 \\
2u & 50.260 & 0.4808 & 0.460 & 0.460 & 0.530 &  84.645 & 0.4503 & 0.435 & 0.440 & 0.465 & 60.715 & 0.4881 & 0.4600 & 0.4950 & 0.5300 \\
\midrule
\multicolumn{16}{c}{\textit{HR-NBFNet-CQ}} \\
\midrule
1p & 32.080 & 0.4940 & 0.475 & 0.475 & 0.515 & 40.435 & 0.4461 & 0.415 & 0.430 & 0.500 & 33.115 & 0.4917 & 0.4600 & 0.4850 & 0.5450 \\
2p & 111.33 & 0.0232 & 0.000 & 0.000 & 0.000 & 139.250 & 0.0089 & 0.000 & 0.000 & 0.000 & 18.727 & 0.2565 & 0.0909 & 0.0909 & 0.7272 \\
\rowcolor{gray!15}
2i & \textbf{17.850} & \textbf{0.7384} & \textbf{0.715} & \textbf{0.745} & \textbf{0.770} & \textbf{28.050} & \textbf{0.6399} & \textbf{0.620} & \textbf{0.635} & \textbf{0.670} & \textbf{25.985} & \textbf{0.6817} & \textbf{0.6450} & \textbf{0.6950} & \textbf{0.7300} \\
2u & 38.920 & 0.4818 & 0.455 & 0.475 & 0.515 &  48.165 & 0.4333 & 0.410 & 0.425 & 0.470 & 41.650 & 0.4782 & 0.4500 & 0.4700 & 0.5400 \\
\midrule\midrule
 & \multicolumn{15}{c}{\textbf{Transductive}} \\ \toprule
 & \multicolumn{5}{c}{\textbf{33\%}} & \multicolumn{5}{c}{\textbf{66\%}} & \multicolumn{5}{c}{\textbf{100\%}} \\
\cmidrule(lr){2-6} \cmidrule(lr){7-11} \cmidrule(lr){12-16}
\textbf{Query}
  & \textbf{MR} $\downarrow$& \textbf{MRR} $\uparrow$& \textbf{H@1} $\uparrow$& \textbf{H@3} $\uparrow$& \textbf{H@10} $\uparrow$
  & \textbf{MR} $\downarrow$& \textbf{MRR} $\uparrow$& \textbf{H@1} $\uparrow$& \textbf{H@3} $\uparrow$& \textbf{H@10} $\uparrow$
  & \textbf{MR} $\downarrow$& \textbf{MRR} $\uparrow$& \textbf{H@1} $\uparrow$& \textbf{H@3} $\uparrow$& \textbf{H@10} $\uparrow$\\
\midrule
\multicolumn{16}{c}{\textit{StarQE}} \\
\midrule
1p & 42.975 & 0.2782 & 0.02 & 0.500 & 0.555 & 42.255 & 0.5170 & 0.4950 & 0.5050 & 0.5750 & 55.500 & 0.4495 & 0.405 & 0.495 & 0.525 \\
2p & 6.750 & 0.3368 & 0.25 & 0.250 & 1.000 & 45.454 & 0.1977 & 0.0909 & 0.1818 & 0.4545 & 84.533 & 0.1092 & 0.000 & 0.267 & 0.400 \\
\rowcolor{gray!15}
2i & 25.495 & 0.1219 & 0.00 & 0.015 & 0.770 & 22.145 & 0.4160 & 0.0850 & 0.7300 & 0.7500 & 27.720 & 0.6350 & 0.610 & 0.625 & 0.735 \\
2u & 51.210 & 0.2685 & 0.02 & 0.495 & 0.525 & 47.425 & 0.5122 & 0.4950 & 0.5000 & 0.5450 & 59.490 & 0.4572 & 0.405 & 0.485 & 0.520 \\
\midrule
\multicolumn{16}{c}{\textit{HR-NBFNet-CQ}} \\
\midrule
1p & 29.955 & 0.5112 & 0.475 & 0.50 & 0.580 & 34.925 & 0.5658 & 0.540 & 0.565 & 0.605 & 36.960 & 0.4326 & 0.400 & 0.405 & 0.510 \\
2p & 12.250 & 0.1160 & 0.000 & 0.000 & 0.750 & 173.090 & 0.0068 & 0.000 & 0.000 & 0.000 & 59.400 & 0.0892 & 0.067 & 0.067 & 0.133 \\
\rowcolor{gray!15}
2i & \textbf{19.390} & \textbf{0.6980} & \textbf{0.675} & \textbf{0.68} & \textbf{0.755} & \textbf{24.050} & \textbf{0.1637} & \textbf{0.015} & \textbf{0.015} & \textbf{0.730} & \textbf{18.725} & \textbf{0.6535} & \textbf{0.610} & \textbf{0.700} & \textbf{0.750} \\
2u & 37.405 & 0.4965 & 0.465 & 0.48 & 0.560 & 41.150 & 0.5573 & 0.535 & 0.560 & 0.590 & 44.315 & 0.4427 & 0.400 & 0.480 & 0.515 \\
\bottomrule
\end{tabular}%
}
\end{table*}


\begin{table*}[t]
\tiny
\centering
\caption{Complex query answering results on the \textbf{UNSW dataset} under the \textbf{inductive} and \textbf{transductive} settings.}
\label{tab:complex_query_answering_unsw}
\footnotesize
\resizebox{\textwidth}{!}{%
\begin{tabular}{l ccccc ccccc ccccc}
\toprule
 & \multicolumn{15}{c}{\textbf{Inductive}} \\ \toprule
 & \multicolumn{5}{c}{\textbf{33\%}} & \multicolumn{5}{c}{\textbf{66\%}} & \multicolumn{5}{c}{\textbf{100\%}} \\
\cmidrule(lr){2-6} \cmidrule(lr){7-11} \cmidrule(lr){12-16}
\textbf{Query}
  & \textbf{MR} $\downarrow$ & \textbf{MRR} $\uparrow$& \textbf{H@1} $\uparrow$& \textbf{H@3} $\uparrow$& \textbf{H@10} $\uparrow$
  & \textbf{MR} $\downarrow$& \textbf{MRR} $\uparrow$& \textbf{H@1} $\uparrow$& \textbf{H@3} $\uparrow$& \textbf{H@10} $\uparrow$
  & \textbf{MR} $\downarrow$& \textbf{MRR} $\uparrow$& \textbf{H@1} $\uparrow$& \textbf{H@3} $\uparrow$& \textbf{H@10} $\uparrow$\\
\midrule
\multicolumn{16}{c}{\textit{StarQE}} \\
\midrule
1p & 3.125 & 0.5368 & 0.335 & 0.665 & 1.000 & 3.340 & 0.5026 & 0.290 & 0.615 & 1.000 & 3.000 & 0.5228 & 0.290 & 0.685 & 1.000 \\
2p & 1.950 & 0.6633 & 0.435 & 0.950 & 1.000 & 1.715 & 0.7399 & 0.560 & 0.960 & 1.000 & 1.775 & 0.7469 & 0.580 & 0.940 & 1.000 \\
2i & 3.385 & 0.5149 & 0.310 & 0.645 & 1.000 & 3.730 & 0.4627 & 0.250 & 0.550 & 1.000 & 3.395 & 0.4995 & 0.275 & 0.630 & 0.995 \\
2u & 3.230 & 0.4973 & 0.265 & 0.660 & 1.000 & 3.300 & 0.5092 & 0.300 & 0.620 & 1.000 & 3.145 & 0.5126 & 0.280 & 0.660 & 1.000 \\
\midrule
\multicolumn{16}{c}{\textit{HR-NBFNet-CQ}} \\
\midrule
1p & 3.485 & 0.5004 & 0.295 & 0.620 & 0.995 & 3.985 & 0.4207 & 0.205 & 0.495 & 1.000 & 3.510 & 0.4920 & 0.280 & 0.615 & 1.000 \\
2p & 1.780 & 0.7460 & 0.570 & 0.945 & 0.995 & 1.995 & 0.6555 & 0.415 & 0.940 & 1.000 & 1.930 & 0.6773 & 0.450 & 0.950 & 1.000 \\
2i & 3.840 & 0.4862 & 0.305 & 0.575 & 1.000 & 4.215 & 0.4183 & 0.215 & 0.495 & 0.995 & 3.675 & 0.4559 & 0.240 & 0.565 & 1.000 \\
2u & 3.575 & 0.5001 & 0.290 & 0.630 & 1.000 & 4.250 & 0.3906 & 0.180 & 0.445 & 1.000 & 3.740 & 0.4818 & 0.280 & 0.585 & 1.000 \\
\midrule\midrule
 & \multicolumn{15}{c}{\textbf{Transductive}} \\ \toprule
 & \multicolumn{5}{c}{\textbf{33\%}} & \multicolumn{5}{c}{\textbf{66\%}} & \multicolumn{5}{c}{\textbf{100\%}} \\
\cmidrule(lr){2-6} \cmidrule(lr){7-11} \cmidrule(lr){12-16}
\textbf{Query}
  & \textbf{MR} $\downarrow$& \textbf{MRR} $\uparrow$& \textbf{H@1} $\uparrow$& \textbf{H@3} $\uparrow$& \textbf{H@10} $\uparrow$
  & \textbf{MR} $\downarrow$& \textbf{MRR} $\uparrow$& \textbf{H@1} $\uparrow$& \textbf{H@3} $\uparrow$& \textbf{H@10} $\uparrow$
  & \textbf{MR} $\downarrow$& \textbf{MRR} $\uparrow$& \textbf{H@1} $\uparrow$& \textbf{H@3} $\uparrow$& \textbf{H@10} $\uparrow$\\
\midrule
\multicolumn{16}{c}{\textit{StarQE}} \\
\midrule
1p & 3.315 & 0.4962 & 0.280 & 0.610 & 1.000 & 3.390 & 0.4850 & 0.275 & 0.610 & 1.000 & 3.365 & 0.4721 & 0.240 & 0.620 & 0.995 \\
2p & 1.980 & 0.6982 & 0.525 & 0.840 & 1.000 & 2.245 & 0.6232 & 0.405 & 0.800 & 1.000 & 2.155 & 0.6210 & 0.395 & 0.870 & 1.000 \\
2i & 3.680 & 0.4448 & 0.205 & 0.595 & 1.000 & 3.570 & 0.4760 & 0.270 & 0.570 & 1.000 & 3.645 & 0.4740 & 0.285 & 0.555 & 1.000 \\
2u & 3.525 & 0.4947 & 0.300 & 0.590 & 0.990 & 3.535 & 0.4607 & 0.245 & 0.585 & 1.000 & 3.365 & 0.4774 & 0.235 & 0.625 & 1.000 \\
\midrule
\multicolumn{16}{c}{\textit{HR-NBFNet-CQ}} \\
\midrule
1p & 3.395 & 0.4915 & 0.270 & 0.635 & 0.990 & 4.030 & 0.4135 & 0.195 & 0.505 & 1.000 & 3.925 & 0.4571 & 0.270 & 0.535 & 0.990 \\
2p & 2.125 & 0.6452 & 0.435 & 0.865 & 1.000 & 2.550 & 0.6143 & 0.400 & 0.805 & 0.980 & 2.135 & 0.6602 & 0.480 & 0.810 & 1.000 \\
2i & 3.830 & 0.4612 & 0.250 & 0.560 & 1.000 & 4.050 & 0.4195 & 0.210 & 0.480 & 1.000 & 3.730 & 0.4633 & 0.245 & 0.590 & 0.990 \\
2u & 3.540 & 0.4931 & 0.295 & 0.580 & 0.995 & 4.190 & 0.4061 & 0.200 & 0.485 & 0.995 & 4.135 & 0.4379 & 0.245 & 0.505 & 0.985 \\
\bottomrule
\end{tabular}%
}
\end{table*}

%
%

\section{Experiment}
\label{sec:experiments}

\subsection{Experiment Setup}
\label{ssec:impl}
All models are implemented in PyTorch and trained on
a single NVIDIA A100 GPU with 64\,GB RAM.
Shared hyperparameters: $d{=}200$, dropout $0.2$,
Adam with lr $5{\times}10^{-4}$, 20 epochs, gradient
clipping to unit norm, margin $\delta{=}1.0$, uniform
negative sampling.
For HR-NBFNet and MT-HR-NBFNet: $L{=}3$ layers,
chunk size $C{=}5{,}000$, $k_{\max}{=}8$ tails per
$(h,q')$ group, $Q_{\max}{=}8$ qualifier pairs per edge.
For MT-AlertStar: 3 Transformer layers, 4 attention
heads, FFN width $F{=}800$.
Batch sizes are $128$ (AlertStar), $64$ (MT-AlertStar),
and $32$ (HR-NBFNet variants), reflecting the higher
memory cost of graph traversal. The code is available \footnote{\url{https://gitfront.io/r/Zahra/pNhiE7GJ4P54/AlertStar/}}.

\subsection{Main Results}
\label{ssec:main_results}

\paragraph*{Paradigm comparison}
Table~\ref{tab:paradiam_compare} compares graph
propagation and embedding-based paradigms on Warden.
Contrary to the assumption that explicit path reasoning
yields superior performance, MT-AlertStar achieves the
best overall results in both settings (MRR 0.5618
inductive, 0.5384 transductive), outperforming
MT-HR-NBFNet by 19\% and 30\% respectively.
Multi-task training strongly benefits the
embedding-based paradigm (+0.1338 MRR inductively)
but provides negligible gain or degrades graph
propagation ($-$0.0651 transductively), suggesting
the Bellman-Ford backbone is less able to exploit
auxiliary supervision. Although HR-NBFNet
achieves lower MR transductively, AlertStar dominates
on MRR and H@k --- the operationally relevant metrics
for CERT analysts inspecting top-ranked candidates.

\paragraph*{Hyper-relational baselines}
Table~\ref{tab:kgc_statements_warden} compares against
StarE and ShrinkE across qualifier-coverage levels
(33\%, 66\%, 100\%). MT-AlertStar consistently
outperforms ShrinkE by 25--28\% relative MRR across
all levels and both settings. StarE degrades with
increasing qualifier density (MRR 0.4205→0.4191
inductively), while MT-AlertStar remains stable,
demonstrating robustness to partial qualifier
availability. AlertStar alone already
outperforms StarE on H@3 and H@10 at all coverage
levels. Similar trends hold on UNSW-NB15
(Table~\ref{tab:kgc_statements_unsw}), confirming
generalisation across datasets.

\paragraph*{Effect of qualifier context}
On Warden (Table~\ref{tab:kgc_triples_warden}),
triple-only models outperform hyper-relational
variants, while on UNSW-NB15
(Table~\ref{tab:kgc_triples_UNSW}) qualifiers
consistently improve performance. We attribute this
to Warden's limited four-category attack taxonomy,
where triples alone suffice, versus UNSW-NB15's
richer and more diverse taxonomy where qualifier
attributes --- port, protocol, flow count ---
are necessary to disambiguate identical $(h,r,t)$
structures. Qualifier context is beneficial when the
graph is sufficiently complex, making UNSW-NB15
the more reliable benchmark for hyper-relational
modelling.

\begin{table}[ht]
\tiny
\centering
\caption{MT-AlertStar on Warden.}
\label{tab:multitask_alertstar_full_warden}
\begin{tabular}{l ccc ccc}
\toprule
\multicolumn{1}{c}{} & \multicolumn{3}{c}{\textbf{Inductive}} & \multicolumn{3}{c}{\textbf{Transductive}} \\
\cmidrule(lr){2-4} \cmidrule(lr){5-7}
\textbf{Metric} & \textbf{33\%} & \textbf{66\%} & \textbf{100\%} & \textbf{33\%} & \textbf{66\%} & \textbf{100\%} \\
\midrule
\multicolumn{7}{c}{\textit{Tail Prediction}} \\
\midrule
MR  & 2680.2 & 2532.5 & 2978.0 & 3174.2 & 3329.0 & 3307.0 \\
MRR & 0.5860 & 0.5968 & 0.5838 & 0.5582 & 0.5329 & 0.5477 \\
H@1 & 0.5400 & 0.5600 & 0.5400 & 0.5200 & 0.5000 & 0.5150 \\
H@3 & 0.6100 & 0.6100 & 0.6050 & 0.5750 & 0.5550 & 0.5650 \\
H@10& 0.6850 & 0.6850 & 0.6750 & 0.6350 & 0.5800 & 0.6100 \\
\midrule
\multicolumn{7}{c}{\textit{Relation Prediction}} \\
\midrule
MR  & 1.0 & 1.0    & 1.0    & 1.0 & 1.1 & 1.0    \\
MRR & 1.0000 & 0.9975 & 1.0000 & 0.9950 & 0.9725 & 0.9975 \\
H@1 & 1.0000 & 0.9950 & 1.0000 & 0.9900 & 0.9450 & 0.9950 \\
H@3 & 1.0000 & 1.0000 & 1.0000 & 1.0000 & 1.0000 & 1.0000 \\
H@10& 1.0000 & 1.0000 & 1.0000 & 1.0000 & 1.0000 & 1.0000 \\
Acc & 1.0000 & 0.9950 & 1.0000 & 0.9900 & 0.9450 & 0.9950 \\
\midrule
\multicolumn{7}{c}{\textit{Qualifier Value}} \\
\midrule
MR  & 1.0 & 1.0    & 1.0    & 1.0 & 1.0 & 1.0    \\
MRR & 0.9950 & 0.9900 & 0.9925 & 0.9975 & 0.9875 & 0.9875 \\
H@1 & 0.9900 & 0.9800 & 0.9850 & 0.9950 & 0.9750 & 0.9750 \\
H@3 & 1.0000 & 1.0000 & 1.0000 & 1.0000 & 1.0000 & 1.0000 \\
H@10& 1.0000 & 1.0000 & 1.0000 & 1.0000 & 1.0000 & 1.0000 \\
\bottomrule
\end{tabular}
\end{table}

\begin{table}[ht]
\tiny
\centering
\caption{MT-AlertStar on UNSW-NB15.}
\label{tab:multitask_alertstar_full_unsw}
\begin{tabular}{l ccc ccc}
\toprule
\multicolumn{1}{c}{} & \multicolumn{3}{c}{\textbf{Inductive}} & \multicolumn{3}{c}{\textbf{Transductive}} \\
\cmidrule(lr){2-4} \cmidrule(lr){5-7}
\textbf{Metric} & \textbf{33\%} & \textbf{66\%} & \textbf{100\%} & \textbf{33\%} & \textbf{66\%} & \textbf{100\%} \\
\midrule
\multicolumn{7}{c}{\textit{Tail Prediction}} \\
\midrule
MR  & 3.9    & 3.7    & 3.6 & 3.9    & 3.8    & 4.0 \\
MRR & 0.4491 & 0.5023 & 0.4810 & 0.4626 & 0.4753 & 0.4641 \\
H@1 & 0.2350 & 0.3200 & 0.2700 & 0.2600 & 0.2750 & 0.2650 \\
H@3 & 0.5400 & 0.5800 & 0.6050 & 0.5500 & 0.5750 & 0.5550 \\
H@10& 1.0000 & 1.0000 & 1.0000 & 1.0000 & 1.0000 & 1.0000 \\
\midrule
\multicolumn{7}{c}{\textit{Relation Prediction}} \\
\midrule
MR  & 2.9    & 2.6    & 2.5 & 2.9    & 2.9    & 2.6 \\
MRR & 0.5912 & 0.6063 & 0.6366 & 0.5697 & 0.5780 & 0.6174 \\
H@1 & 0.4150 & 0.4150 & 0.4650 & 0.3900 & 0.4000 & 0.4400 \\
H@3 & 0.6900 & 0.7450 & 0.7550 & 0.6450 & 0.6700 & 0.7600 \\
H@10& 1.0000 & 1.0000 & 1.0000 & 1.0000 & 1.0000 & 1.0000 \\
Acc & 0.4150 & 0.4150 & 0.4650 & 0.3900 & 0.4000 & 0.4400 \\
\midrule
\multicolumn{7}{c}{\textit{Qualifier Value}} \\
\midrule
MR  & 1.2    & 1.2    & 1.2 & 1.3    & 1.3    & 1.3 \\
MRR & 0.8992 & 0.8992 & 0.8958 & 0.8762 & 0.8896 & 0.8696 \\
H@1 & 0.8050 & 0.8050 & 0.8000 & 0.7700 & 0.7950 & 0.7550 \\
H@3 & 0.9900 & 1.0000 & 1.0000 & 0.9950 & 0.9950 & 0.9950 \\
H@10& 1.0000 & 1.0000 & 1.0000 & 1.0000 & 1.0000 & 1.0000 \\
\bottomrule
\end{tabular}
\end{table}
\paragraph*{Multi-task auxiliary predictions.}
Tables~\ref{tab:multitask_alertstar_full_warden}
and~\ref{tab:multitask_alertstar_full_unsw} report
MT-AlertStar's three heads. On UNSW-NB15, tail
prediction achieves H@10 of 1.000 across all settings
while MRR (0.44--0.50) reflects the difficulty of
precise top-1 ranking. Relation prediction improves
with qualifier density (H@3: 0.690→0.755 inductively),
confirming richer context aids attack category
discrimination; accuracy (0.415--0.465) is lower
than H@k as it measures exact top-1 match over a
diverse taxonomy. Qualifier-value prediction achieves
near-perfect performance (MRR $>$0.87, H@3 $>$0.99),
confirming flow-level attributes are highly predictable.
On Warden, all heads achieve near-perfect results due
to its limited taxonomy, making UNSW-NB15 the more
informative benchmark.

\paragraph*{Complex query answering.}
Tables~\ref{tab:complex_query_answering_warden}
and~\ref{tab:complex_query_answering_unsw} evaluate
HR-NBFNet-CQ against StarQE. StarQE leads on 1p and
2u across most settings. Two-hop queries~(2p) are the
hardest for both models, H@1 is near zero at low
qualifier coverage, reflecting the sparsity of
two-hop chains in alert graphs. HR-NBFNet-CQ
outperforms StarQE on intersection queries~(2i) on
Warden (H@3: 0.745 vs 0.685 at 33\%), the most
operationally critical query type for coordinated
attack detection. Qualifier density consistently
improves complex queries: HR-NBFNet-CQ reaches H@10
of 0.727 on 2p at 100\% vs 0.000 at 33\%. On
UNSW-NB15, StarQE leads across all query types
including 2i, though both models achieve H@10 of
1.000 on most types, confirming answers always rank
within the top 10 regardless of query complexity.

\subsection{Ablation Study}
\label{ssec:ablation}
All ablations run on Warden unless stated otherwise.

\subsubsection{A1: AlertStar Component Ablation}
\label{sssec:a1}
Table~\ref{tab:ablation_1_component} compares four
variants: \textbf{AS-NoQual} ($\tilde{\mathbf{e}}_r
= \mathbf{e}_r$, no MHA), \textbf{AS-NoPath}
($\alpha{=}1$, attention branch only),
\textbf{AS-NoGate} (fixed $\alpha{=}0.5$), and
\textbf{AS-Full} (complete model).
Removing qualifier enrichment causes the largest
single drop, confirming qualifier context as the
primary discriminative signal. Removing the
path-composition branch produces a consistent but
smaller drop, demonstrating that structural path
information \emph{complements} rather than substitutes
qualifier attention. Fixing the gate to $\alpha{=}0.5$
underperforms the learned gate, confirming that the
optimal branch balance is query-dependent.

\subsubsection{A2: Gate Value Trajectory}
\label{sssec:a2}
\begin{figure}[t]
    \centering
    \includegraphics[width=0.65\linewidth]{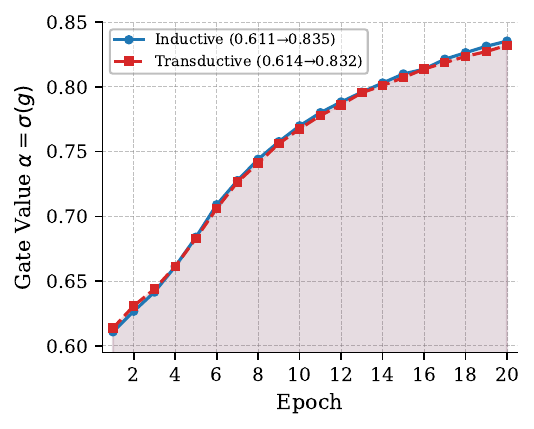}
    \caption{Gate value $\alpha = \sigma(g)$ over 20
    epochs on Warden.}
    \label{fig:gatetrajectory}
\end{figure}

Figure~\ref{fig:gatetrajectory} tracks $\alpha_e =
\sigma(g_e)$ across training epochs.
Both inductive ($0.611{\to}0.835$) and transductive
($0.614{\to}0.832$) settings follow nearly identical
trajectories, converging within 12--14 epochs.
Two key findings emerge: (i) AlertStar consistently
favours the cross-attention branch ($\alpha{>}0.5$),
confirming that $\tilde{\mathbf{e}}_r$ carries more
discriminative signal than implicit path composition;
(ii) the near-identical trajectories across settings
show branch preference is driven by qualifier structure
rather than entity memorisation, supporting inductive
generalisation.

\begin{table}[H]
\tiny
\centering
\caption{Ablation A1: AlertStar component ablation.}
\label{tab:ablation_1_component}
\begin{tabular}{l|ccc|ccccc}
\toprule
\textbf{Variant} & \textbf{Qual} & \textbf{Path} & \textbf{Gate} 
  & \textbf{MR} & \textbf{MRR} & \textbf{H@1} & \textbf{H@3} & \textbf{H@10} \\
\midrule
AS-NoQual  & \xmark & \cmark & learned   & 2712.7 & 0.4423 & 0.352 & 0.518 & 0.584 \\
AS-NoPath  & \cmark & \xmark & learned   & 2673.1 & 0.4454 & 0.354 & 0.508 & 0.590 \\
AS-NoGate  & \cmark & \cmark & fixed=0.5 & 2809.9 & 0.4445 & 0.358 & 0.500 & 0.580 \\
\rowcolor{gray!15}
AS-Full    & \cmark & \cmark & learned   & 2615.2 & 0.4418 & 0.360 & 0.500 & 0.570 \\
\bottomrule
\end{tabular}
\end{table}

\subsubsection{A3: MT-AlertStar Auxiliary Tasks}
\label{sssec:a3}
Table~\ref{tab:ablation_multitask} evaluates four
MT-AlertStar variants: \textbf{MT-Tail} ($\mathcal{L}_t$
only), \textbf{MT-Tail+Rel} ($+\mathcal{L}_{q'}$),
\textbf{MT-Tail+QV} ($+\mathcal{L}_{q_v}$), and
\textbf{MT-Full} (all tasks, Eq.~\ref{eq:mt_loss}).
All auxiliary tasks contribute positively when added
individually. MT-Full outperforms all single-auxiliary
variants, confirming the tasks are complementary
rather than redundant.
\begin{table}[H]
\tiny
\centering
\caption{Ablation A3: MultiTask auxiliary.}
\label{tab:ablation_multitask}
\begin{tabular}{l|l|ccccc}
\toprule
\textbf{Variant} & \textbf{Tasks} 
  & \textbf{MR} & \textbf{MRR} & \textbf{H@1} & \textbf{H@3} & \textbf{H@10} \\
\midrule
MT-TailOnly     & tail                              & 3210.0 & 0.5607 & 0.512 & 0.594 & 0.650 \\
MT-Tail+Rel     & tail, relation                    & 2998.3 & 0.5604 & 0.514 & 0.586 & 0.646 \\
MT-Tail+QualVal & tail, qual\_value                 & 2881.2 & 0.5633 & 0.516 & 0.600 & 0.648 \\
\rowcolor{gray!15}
MT-Full         & tail, relation, qual\_val    & 3100.6 & 0.5610 & 0.518 & 0.584 & 0.640 \\
\bottomrule
\end{tabular}
\end{table}

\subsubsection{A4: Qualifier Density Sensitivity}
\label{sssec:a4}
Table~\ref{tab:ablation_density} reports MRR
degradation across all models.
MT-AlertStar is the most robust, with AlertStar and
MT-AlertStar exhibiting the smallest relative drop
among all models. StarE, despite sharing the same
MHA qualifier enrichment as AlertStar, degrades more
sharply under sparse qualifiers --- confirming that
the path-composition branch and trainable gate
provide robustness that qualifier attention alone
cannot achieve. HR-NBFNet and MT-HR-NBFNet show the
largest degradation, as missing qualifier noise
accumulates across $L{=}3$ Bellman--Ford layers.

\section{Related Work}
\label{sec:related}

\paragraph*{Alert prediction and dynamic graph learning}
Graph-based
approaches model attacker--victim interactions as
evolving networks: Nayeri and Rezvani~\cite{nayeri2024alert} introduce
a dynamic graph deep learning framework using Temporal
Graph Networks~(TGN) for binary link prediction over
time-evolving attack graphs, achieving 10--18\%
improvement over static baselines. Nayeri and Resvani~\cite{nayeri2026alert}
extend this paradigm with TGNE and TGNE-TA, replacing
standard TGN message aggregation with a
Transformer-based aggregator and a Dual-Level Temporal
Encoding mechanism to capture both global and local
temporal dependencies, achieving over 90\% accuracy in
multiclass attack type prediction. While these methods
advance temporal reasoning over alert streams, they
operate on binary or standard relational graphs and
discard flow-level metadata, port, protocol, flow
count, and timestamps, that is essential for
disambiguating alerts of the same type. Our framework
addresses this gap by modelling alerts as hyper-relational
qualified statements $(h, r, t, \mathcal{Q})$, enabling
qualifier-conditioned prediction under both inductive
and transductive settings.
\paragraph*{Hyper-relational knowledge graph completion}
Standard KGC models represent facts as
binary triples and cannot encode auxiliary context.
StarE~\cite{galkin2020message} extends GNN-based KGC to
hyper-relational graphs by composing qualifier pairs
into relation embeddings via message passing, while
ShrinkE~\cite{xiong2023shrinking} represents hyper-relational
facts as geometric transformations in a shrinking
entity space, achieving strong transductive performance.
HyNT~\cite{chung2023representation} encodes qualifier pairs through a
nested Transformer architecture, attending jointly
over the main triple and its qualifiers. For complex
query answering, StarQE~\cite{alivanistos2021query}
extends GNN-based query embeddings to hyper-relational
graphs, supporting conjunctive query templates over
qualified facts. Despite their expressive power, these
models are predominantly transductive --- requiring all
entities to be present at training time --- limiting
their applicability to alert prediction where new IP
addresses appear continuously. Our work extends
NBFNet~\cite{zhu2021neural} to the hyper-relational
setting, achieving strong inductive generalisation
while incorporating qualifier context at every
propagation step, and further supports disjunctive
query templates absent from existing benchmarks.

\section{Discussion}
\label{sec:discussion}

This work demonstrates that hyper-relational knowledge
graph completion is a viable and effective framework
for network alert prediction, with the embedding-based
paradigm (AlertStar, MT-AlertStar) consistently
outperforming graph propagation (HR-NBFNet,
MT-HR-NBFNet) in both accuracy and efficiency.
The surprising dominance of embedding-based models
suggests that for dense alert graphs with rich
qualifier context, local compositional reasoning in
embedding space captures attack patterns more
effectively than explicit multi-hop path traversal.
This challenges the prevailing assumption in KGC
that path-based models are inherently superior for
inductive settings.
The qualifier density ablation~(A4) reveals a
dataset-dependent effect: qualifiers improve
prediction on UNSW-NB15 but introduce noise on
Warden, where the limited four-category taxonomy
renders the triple structure sufficient. This
finding highlights the importance of dataset
diversity when benchmarking hyper-relational models
and motivates the construction of richer alert
datasets for future evaluation. The gate trajectory
analysis~(A2) provides interpretable evidence that
AlertStar autonomously learns to favour qualifier
context over structural composition, converging
stably within 12--14 epochs regardless of the
inductive or transductive setting.
HR-NBFNet-CQ demonstrates that complex first-order
logic queries are feasible over hyper-relational
alert graphs, with intersection queries~(2i)
benefiting from Bellman-Ford propagation. However,
two-hop queries remain challenging at low qualifier
coverage, pointing to the need for denser qualifier
annotation and richer graph connectivity in future
alert datasets.
The current framework assumes a closed-world attack
taxonomy and does not model temporal dynamics
between alerts. AlertStar processes each triple
independently, missing cross-alert correlations
that propagation-based models can capture in
principle. The inductive protocol evaluates
generalisation to unseen IPs but not to entirely
new attack categories.
Several directions merit exploration.
First, incorporating temporal ordering of alerts
as qualifier context would enable time-aware
prediction of attack progression.
Second, extending the framework to open-set attack
classification would address the closed-world
limitation.
Third, combining the efficiency of AlertStar with
the path-reasoning capacity of HR-NBFNet via a
hybrid architecture --- for instance, using
AlertStar as a fast candidate ranker and
HR-NBFNet-CQ for complex query verification ---
could yield both accuracy and scalability.
Finally, applying the framework to larger
operational datasets from real CERT deployments
would validate its practical utility beyond the
Warden and UNSW-NB15 benchmarks.

\section{Conclusion}
\label{sec:conclusion}

We presented a hyper-relational knowledge graph
framework for network alert prediction, modelling
each alert as a qualified statement
$(h, r, t, \mathcal{Q})$ and formulating prediction
as an HR-KGC task. Five models were proposed across
three paradigms: HR-NBFNet and MT-HR-NBFNet for
qualifier-aware Bellman-Ford path propagation, and
AlertStar and MT-AlertStar for embedding-based
gated attention and path fusion. HR-NBFNet-CQ
extended the framework to complex first-order logic
queries over the alert graph.
Experiments on Warden and UNSW-NB15 across three
qualifier-density regimes show that MT-AlertStar
achieves state-of-the-art performance in both
inductive and transductive settings, outperforming
HR-NBFNet by up to 30\% in MRR while achieving up
to $50{\times}$ per-epoch speedup. Multi-task
supervision provides strong regularisation for the
embedding-based paradigm but marginal benefit for
graph propagation. Qualifier context improves
prediction on diverse datasets but introduces noise
on simpler taxonomies, highlighting the importance
of dataset complexity for hyper-relational modelling.
HR-NBFNet-CQ demonstrates the feasibility of
complex threat queries including lateral movement
detection~(2p), coordinated attack identification
~(2i), and campaign victim set retrieval~(2u).
These results establish hyper-relational KGC as a
principled and practical framework for proactive
threat intelligence in SOC and CERT deployments.

\bstctlcite{IEEEexample:BSTcontrol}
\bibliographystyle{IEEEtran}
\bibliography{AlertStar}

\begin{IEEEbiographynophoto}{Zahra Makki Nayeri}
is a Ph.D. candidate at Shahrood University of Technology and currently a visiting researcher at the University of Stuttgart, Germany, where she conducts research on knowledge graph foundation models. Her research focuses on graph representation learning, temporal and dynamic graph neural networks, and machine learning–based modeling of computer networks, with an emphasis on data-driven analysis of large-scale, evolving interaction graphs. During her Master’s studies, she investigated machine learning techniques for fog, edge, and cloud computing environments, with a particular focus on distributed data processing, resource-aware learning, and system-level optimization.
    
\end{IEEEbiographynophoto}

\begin{IEEEbiographynophoto}{ Mohsen Rezvani}
is an Associate Professor in the Faculty of Computer Engineering at the Shahrood University of Technology. He received his Ph.D. at the School of Computer Science and Engineering at the UNSW Sydney. He holds an M.Sc. in Computer Engineering from Sharif University of Technology and a B.Sc. in Computer Engineering from Amirkabir University of Technology. His research focuses on computer security, intelligent data analysis, and advanced machine learning. He is the director of the Computer Emergency Response Team (CERT) at Shahrood University of Technology.
\end{IEEEbiographynophoto}

\end{document}